\documentclass[12pt]{iopart} % use larger type; default would be 10pt

\usepackage{iopams,amsfonts,amssymb,amsthm} 

\pdfoutput=1
\usepackage{etex}
\usepackage{cite}
\usepackage{latexsym}
\usepackage{rawfonts}
\usepackage{graphicx}
\usepackage[usenames,dvipsnames,svgnames]{xcolor}
\usepackage{bbm}
\usepackage{latexsym}
\usepackage{multirow}
\usepackage{rotating}
\usepackage{lscape}
\usepackage{graphicx} %\graphicspath{{figs/}}
\usepackage{subfigure}
\usepackage{xcolor}
\usepackage{fancybox}
\usepackage{ifmtarg}

\input prepictex
\input pictex
\input postpictex

\usepackage{cmbright}
\usepackage[T1]{fontenc}
\usepackage{graphicx}
\def\q{\quad}
\def\qq{\qquad}

\def\2;{\;\;}

\def\eps{\epsilon}
\def\Pr{{\hbox{P}_{\hskip -1mm r}}}

\def\mathL{{\mathbb L}}

\def\shalf{{\sfrac{1}{2}}}

\def\Ref#1{(\ref{#1})}

\def\C#1{{\mathcal #1}}

\def\taus#1{{\tau_#1}}

\def\Sfrac#1#2{\hbox{\large $\frac{#1}{#2}$}}
\def\sfrac#1#2{\hbox{\nor $\frac{#1}{#2}$}}

\def\LB{\left(}         \def\RB{\right)}
        
\def\LA{\left\langle}        \def\RA{\right\rangle}

\def\lfl{\left\lfloor} \def\rfl{\right\rfloor}
       
\def\LH{\left[}        \def\RH{\right]}

\def\nor{\normalsize}
\def\fns{\scriptsize}

\def\alsu{\hbox{almost surely}}

\def\vv{{\;\hbox{\Large $|$}\;}}

\def\edge#1#2{{\langle #1 \hspace{0.85pt}{\sim}\hspace{0.85pt} #2 \rangle}}

\def\thin{ {\hspace{0.75pt}} }
\def\plus{{\hspace{0.85pt}{+}\hspace{0.85pt}}}
\def\minus{{\hspace{0.85pt}{-}\hspace{0.85pt}}}

\usepackage{cmbright}
\usepackage{textcomp}
\usepackage[font=footnotesize,labelfont=sf]{caption}
\definecolor{blue}{rgb}{0,0.18,0.39}
\definecolor{RoyalBlue}{rgb}{0,0.2,0.7}

\begin{document}
\title{Microcanonical Simulations of Adsorbing Self-Avoiding Walks}
\author{
E.J. Janse van Rensburg$^1$\footnote[1]{\texttt{rensburg@yorku.ca}}}

\address{$^1$Department of Mathematics and Statistics, 
York University, Toronto, Ontario M3J~1P3, Canada\\}

\begin{abstract}
Linear polymers adsorbing on a wall can be modelled by half-space
self-avoiding walks adsorbing on a line in the square lattice, or on a 
surface in the cubic lattice.  In this paper a numerical approach based
on the GAS algorithm is used to approximately enumerate states in
the partition function of this model.  The data are used to approximate
the free energy in the model, from which estimates of the location
of the critical point and crossover exponents are made.  The critical
point is found to be located at 
\begin{equation}
a_c^+ =
\cases{
1.779 \pm 0.003, & \hbox{in the square lattice}; \\
1.306 \pm 0.007, & \hbox{in the cubic lattice}.
}
\end{equation} 
These results are then used to estimate the crossover exponent $\phi$
associated with the adsorption transition,  giving
\begin{equation}
\phi =
\cases{
0.496 \pm 0.009, & \hbox{in two dimensions}; \\
0.505 \pm 0.006, & \hbox{in three dimensions}.
}
\end{equation} 
In addition, the scaling of these thermodynamic quantities is examined
using the numerical data, including the scaling of metric quantities, 
and the partition and generating functions.  In all cases results and
numerical values of exponents were obtained which are consistent with
estimates in the literature.
\end{abstract}

%Uncomment for PACS numbers title message
\pacs{82.35.Lr,82.35.Gh,61.25.Hq}
\ams{82B41, 82B80, 65C05}
% Uncomment for Submitted to journal title message
\submitto{J. Stat. Mech. Theor. Exp.}
% Comment out if separate title page not required
\maketitle

\section{Introduction}
\label{section1}   %%%%%%ZXZ[section1]

The adsorption of a linear polymer on an attractive surface is a
conformational rearrangement of the polymer to a state where it 
explores conformations which remain near or on the surface.  
This is the so-called polymer adsorption transition, and the many models
of this phenomenon (see, for example, reference \cite{JvR15}) remain a rich
source of mathematical and numerical studies.  Polymer adsorption is 
a phase transition \cite{BDG83}, and the properties of the adsorbed polymer 
have been examined both experimentally (see, for example, references
\cite{DR71,BDD95,MBM14}) and theoretically \cite{deG79}.  These models 
include directed path models of adsorbing polymers \cite{PFF88,W98}, as well as
self-avoiding walk models \cite{HTW82}, and these have received considerable
attention in the literature \cite{HG94}, using both rigorous methods \cite{RW11}
and numerical methods (for example, the Monte Carlo simulation of adsorbing
self-avoiding walks \cite{JvRR04}).

In this study, a new Monte Carlo method for sampling adsorbing self-avoiding
walks is proposed and implemented.  In particular, the GAS algorithm \cite{JvRR09}
is generalised to sample adsorbing walks in the microcanonical ensemble,
and the data obtained are analysed to estimate the locations of
critical points, and the values of critical exponents and scaling of 
adsorbing walks, in the square and cubic lattices.  The algorithm is related
to the Rosenbluth algorithm \cite{RR55} and to the GARM algoritm \cite{RJvR08}.
The GARM algorithm is related to the PERM algorithm \cite{G97,HG11}, and flat
histogram implementations of PERM \cite{PK04} have been used to sample
states from a flat histogram over state space in a variety of different models
of interacting walks, including collapsing walks \cite{PO00} and adsorbing
walks \cite{KPOL04}. 

\begin{figure}[h]
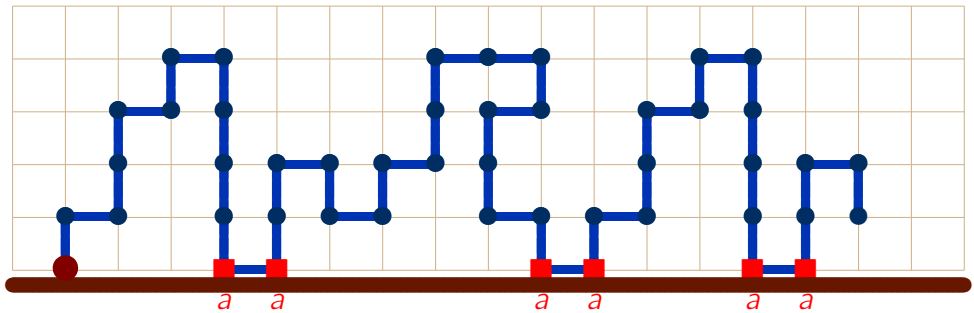

\input figure1-.tex
\caption{An adsorbing self-avoiding walk in the half square lattice 
$\mathL^2_+$.  The walk steps from the origin in the lattice, 
has length $46$, and makes $6$ visits to the adsorbing line
$\partial\mathL^2_+$ (the boundary of $\mathL^2_+$).  The weight
of this walk in the partition function $Z_{46}(a)$ is $a^6$.}
\label{figure0}  %%ZXZ[figure0]
\end{figure}

The GAS algorithm was introduced in reference \cite{JvRR09} and was used 
in several studies as an algorithm to approximately enumerate walks or polygons
\cite{JvRR11,JvRR11A}.  However, it was not clear how to generalise the algorithm to sample states 
in models of interacting walks.  In this paper our purpose is (1) to generalise the
algorithm to a model of adsorbing walks, (2) to examine the behaviour of the
algorithm by computing critical points and exponents of the walk, and to compare this
to results found elsewhere, and (3) to use our data to examine scaling in adsorbing
walks by computing critical exponents and examining the scaling of thermodynamic
functions.

\subsection{Adsorbing walks}
\label{subsection1.2}

An adsorbing self-avoiding walk in the square lattice is illustrated in figure
\ref{figure0}.  Let $\mathL^d$ denote the $d$-dimensional hypercubic lattice
and denote a unit length edge with endpoints $\vec{x}$ and $\vec{y}$ in 
$\mathL^d$ by $\edge{\vec{x}}{\vec{y}}$.  The hypercubic half-lattice 
$\mathL^d_+$ is defined by
\begin{equation}
\mathL^d_+ = 
\{ \edge{\vec{x}}{\vec{y}} \in \mathL^d \vv 
\hbox{$\vec{x}(d) \geq 0$ and $\vec{y}(d) \geq 0$} \},
\end{equation} 
where $\vec{x}(d)$ is the $d$-th Cartesian component of $\vec{x}$, and 
$\vec{y}(d)$ is the $d$-th Cartesian component of $\vec{y}$.   The boundary
of $\mathL^d_+$ is given by
\begin{equation}
\partial\mathL^d_+ = 
\{ \edge{\vec{x}}{\vec{y}} \in \mathL^d_+ \vv 
\hbox{$\vec{x}(d)=0$ and $\vec{y}(d)=0$} \},
\end{equation}
and it is isomorphic to $\mathL^{d-1}$ if $d\geq 2$.   Please note that these definitions,
and the definitions of additional functions and quantities, are listed in table \ref{Defs}.

The number of self-avoiding walks of length $n$ from the origin in $\mathL^d$ is
denoted by $c_n$.  The \textit{growth constant} of the self-avoiding walk 
\cite{H57,H60} is defined by the limit
\begin{equation}
\lim_{n\to\infty} c_n^{1/n} = \mu_d,
\end{equation}
and $\kappa_d = \log \mu_d$ is the connective constant of the self-avoiding walk.
This shows that $c_n = \mu_d^{n+o(n)}$.
Self-avoiding walks from the origin in $\mathL^d_+$ are 
\textit{positive walks}.   The number of positive walks of length $n$
is denoted by $c_n^+$, and it is known that $c_n^+ = \mu_d^{n+o(n)}$ \cite{HTW82}.  

If positive walks are counted with
respect to the number of vertices in $\partial\mathL^d_+$ (these are
\textit{visits}), then the walks are \textit{adsorbing walks}.  For example,
the walk in figure \ref{figure0} is an adsorbing walk with $6$ visits.

Let $c_n^+(v)$ be the number of adsorbing walks of length $n$ from the origin in
$\mathL^d_+$, with $v$ visits to $\partial \mathL^d_+$.  The canonical
partition function of adsorbing walks is obtained by introducing an
activity $a$ conjugate to the number of visits:
\begin{equation}
Z_n(a) = \sum_{v=0}^n c_n^+(v)\,a^v  .
\label{eqnZ}   %%ZXZ[eqnZ]
\end{equation}
When $a$ is large, then $Z_n(a)$ is dominated by walks with a large
number of visits, and if $a$ is small (but positive), then $Z_n(a)$ is 
dominated by walks with a small number of visits.

The \textit{finite size free energy} of these models is computed from the
partition function $Z_n(a)$ (see equation \Ref{eqnZ}), and is given by 
\begin{equation}
\C{F}_n(a) = \Sfrac{1}{n} \log \sum_v c_n^+(v)\, a^v .
\label{eqn23}   %%ZXZ[eqn23]
\end{equation}
The \textit{limiting free energy} of the model is given by the thermodynamic 
limit in the model:
\begin{equation}
\C{F}(a) = \lim_{n\to\infty} \Sfrac{1}{n} \log \sum_v c_n^+(v)\, a^v .
\label{eqn25F}  %%ZXZ[eqn25F]
\end{equation}
This limit exists (see reference \cite{HTW82}, and also, for example, 
reference {\cite{JvR15}}), and it is a convex function of $\log a$ with a 
singular point at $a=a_c^+$ (which is the \textit{adsorption critical point} 
in the model).  For $a<a_c^+$ the model is in a \textit{desorbed state}, 
and for $a>a_c^+$ the model is in an \textit{adsorbed state}.  It is
known that $a_c^+>1$ \cite{JvR98}, and $a_c^+ < \sfrac{\mu_d}{\mu_{d-1}}$
\cite{HTW82}, and 
\begin{equation}
\C{F}(a) \cases{
= \log \mu_d, & if $a\leq a_c^+$; \\
> \log \mu_d, & if $a>a_c^+$.
}
\end{equation}
Since $\C{F}(a)$ is a convex function of $\log a$, it is differentiable for almost all
$a>0$, and it follows that the density of visits to the adsorbing plane is
$a\sfrac{d}{da}\C{F}(a) = 0$ if $a<a_c^+$ (this is the 
\textit{desorbed phase}), and $a\sfrac{d}{da} \C{F}(a) >0$ if $a>a_c^+$ $\alsu$ 
(whenever $\C{F}(a)$ is differentiable).  This is the \textit{adsorbed phase}.
In the desorbed phase the walk tends to make few returns to the adsorbing plane
(walks of length $n$ will return, on average, $o(n)$ times to the adsorbing plane
in the desorbed phase).  Thus, a desorbed walk will tend to move away from the
boundary into the bulk of $\mathL^d_+$.  An adsorbed walk, on the other hand, is expected
to have a positive density of returns to the adsorbing plane.  This implies that the
walk will remain near the adsorbing plane, and so have the properties of a walk
which is stretched out in $(d\minus 1)$ dimensions near $\partial\mathL^d_+$
(and compressed in the $d$-th dimension).  Separating these two regimes is
the adsorption critical point $a_c^+$.  

The finite size scaling of the free energy $\C{F}_n(a)$ is given by 
\begin{equation}
\C{F}_n(a) \simeq \log \mu_d + (a\minus a_c^+)^{2-\alpha}\,f(n^\phi(a\minus a_c^+)),
\label{eqn24}  %%ZXZ[eqn24]
\end{equation}
where $f$ is a scaling function, $\alpha$ is the specific heat
exponent, and $\phi$ is the finite size crossover exponent.   The exponents $\alpha$ 
and $\phi$ are related by the hyperscaling relation
\begin{equation}
2-\alpha=\Sfrac{1}{\phi} .
\label{eqn25}  %%ZXZ[eqn25]
\end{equation}

The bulk
entropy contribution to $\C{F}_n(a)$ is $\log\mu_d$ in $\mathL^d_+$
(where $\mu_d$ is the growth constant of self-avoiding walks in
the square lattice).  Slightly redefining the scaling function, it is found that
\begin{equation}
\C{F}_n(a) \simeq \log \mu_d + \Sfrac{1}{n}\, g(n^\phi (a\minus a_c^+)).
\label{eqn26}  %%ZXZ[eqn26]
\end{equation}
By plotting $n(\C{F}_n(a) \minus \log\mu_d)$ against
$n^\phi(a\minus a_c^+)$, the function $g$ can be uncovered (for $n$ large
and $|a\minus a_c^+|$ small). 

Taking derivatives of $\C{F}_n(a)$ to $\log a$ gives the energy (density) 
$\C{E}_n(a)$ and specific heat $\C{C}_n(a)$ of the model.  The scaling of these 
quantities follows directly from equation \Ref{eqn26}:
\begin{equation}
\C{E}_n(a) \simeq n^{\phi-1}\,h_e(n^\phi(a\minus a_c^+)),
\q \hbox{and}\q
\C{C}_n(a) \simeq n^{\alpha\phi}\,h_c(n^\phi(a\minus a_c^+)),
\label{eqn24EC}  %%ZXZ[eqn24EC]
\end{equation}
for some scaling functions $h_e$ and $h_c$.  In the limit as $n\to\infty$,
$\C{E}_n(a) \to \C{E}(a)$ (the limiting energy density) and 
$\C{C}_n(a)\to\C{C}(a)$ (the limiting specific heat).  Existence of these limits (almost everywhere)
is a consequence of the convexity properties of the limiting free energy
(see for example reference \cite{JvR15}).   Physically, $\C{E}(a)$ is the
density of visits per unit length, and $\C{C}(a)$ is the rate of change
in $\C{E}(a)$ as a function of changes in $\log a$ (it has a maximum at
$a_c^+$).

For adsorbing walks it is thought that $\phi=\sfrac{1}{2}$ in all 
dimensions $d\geq 2$ \cite{BEG89,BY95}, and numerical evidence supporting 
this in dimensions lower than $d=4$ (the upper critical dimension) are
available in references \cite{LM88A,ML88A,BWO99,JvRR04,KPHMBS13}.
If $\phi = \sfrac{1}{2}$, then $\alpha = 0$, so, for example, the
specific heat has scaling $\C{C}_n(a) = h_c(n^\phi(a\minus a_c^+))$, 
and plotting measurements of $\C{C}_n(a)$ against the rescaled
variable $\tau=n^\phi(a\minus a_c^+)$ for small values of $\tau$
should collapse the curves
to a limiting curve (with some finite size corrections to scaling),
exposing the scaling function $h_c$.

The partition function has a more complicated scaling law
(see, for example, equation (23) in reference \cite{JvRR04}, or
section 4.2.2 in reference \cite{JvR15}).  In the high temperature
(or small $a$) regime, the partition function scales as
\begin{equation}
Z_n(a) \simeq B_\lambda n^{\gamma_t -1} h_\lambda(n^\phi |t|)\,
\kappa_-^{n\,|t|^{1/\phi}},\q
\hbox{if $a< a_c^+$},
\label{eqn3}   %%ZXZ[eqn3]
\end{equation}
where $t = (a\minus a_c^+)$ and $\lambda$ denotes the high
temperature regime, and where 
$h_\lambda(x) \simeq |x|^{(\gamma_1-\gamma_t)/\phi}$ and
$\log \kappa_- \simeq |a\minus a_c^+|^{-1/\phi}\log \mu_d$.  
Putting $a=1$, for example, and adsorbing constants and functions of $t$ 
into $B_\lambda$, give 
\begin{equation}
Z_n(1) \simeq B_\lambda n^{\gamma_1-1} \mu_d^n, 
\q\hbox{since $\C{F}(a) = \log \mu_d$ if $a<a_c^+$},
\end{equation}
and the exponent $\gamma_1$ is the entropic exponent
of half-space walks, namely $c_n^+ \sim n^{\gamma_1-1} \mu_d^n$.

At the critical adsorption point $a_c^+$, the above scaling is modified to
\begin{equation}
Z_n(a_c^+) \simeq B_c n^{\gamma_t -1}\,\mu_d^n ,\q
\hbox{if $a= a_c^+$},
\label{eqn15}  %%ZXZ[eqn15]
\end{equation}
where $\gamma_t$ is the entropic exponent associated with adsorbing
walks at the critical adsorption point.  The ensemble of half-space walks at 
the critical point has associated \textit{surface entropic exponent} 
$\gamma_s$, and this is related to $\gamma_t$ by $\gamma_t
= \gamma_s$ (see for example section 9.1.3 in reference \cite{JvR15}).

The scaling in the adsorbed phase is similar to the above, but now
with different exponents
\begin{equation}
Z_n(a) \simeq B_{\tau_0} n^{\gamma_t -1} h_{\taus0} (n^\phi |t|)\,
\kappa_+^{n\,|t|^{1/\phi}},\q
\hbox{if $a> a_c^+$},
\end{equation}
where $h_{\taus0} (x) \simeq |x|^{(\gamma_+ -\gamma_t)/\phi}$,
and where the subscript $\taus0$ denotes  the low temperature
(and large $a$) scaling.
Since $\sfrac{1}{n} \log Z_n(a) = \C{F}(a)\,(1\plus o(1))$, it follows
that $\log \kappa_+  \simeq |a\minus a_c^+|^{-1/\phi} \C{F}(a)$.
This, in particular, gives the scaling
\begin{equation}
\C{F}_s(a) \sim |a \minus a_c^+|^{1/\phi}
\end{equation}
for the singular part of the free energy in the adsorbed phase
(consistent with the hyperscaling relation \Ref{eqn25}).
The scaling of $Z_n(a)$ simplifies here to
\begin{equation}
Z_n(a) \simeq B_{\tau_0} n^{\gamma_+-1} \; e^{n\thin\C{F}(a)} ,\q
\hbox{if $a>a_c^+$}.
\end{equation}
The exponent $\gamma_+$ should be that of adsorbed walks, and so
given by $\gamma_+ = \gamma^{(d-1)}$, the entropic exponent of 
walks in one dimension lower.

{\renewcommand{\baselinestretch}{1.2}
% Definitions
\begin{table}[t!]
\begin{center}
{
\caption{Short list of definitions}
\label{Defs}  %%ZXZ[Defs}]
}
{
\begin{tabular}{l|l}
\hline\hline
Function & Definition \\
\hline
$\mathL^d$ & The hypercubic lattice \\
$\mathL^d_+$ & The half-hypercubic lattice \\
$\partial\mathL^d_+$ & The boundary of $\mathL^d_+$ (it is isomorphic is $\mathL^{d-1}$) \\
$\mu_d$ & The growth constant of self-avoiding walks in $\mathL^d$ \\
$c_n^+(v)$ & The number of positive walks of length $n$ and $v$ visits from $\vec{0}$ in $\mathL^d_+$ \\
$Z_n(a)$  & Partition function of positive adsorbing walks of length $n$ and activity $a$ \\
$\C{F}_n(a)$ & Finite size free energy: $\displaystyle F_n(a) = \sfrac{1}{n} \log Z_n(a)$ \\
$\C{F}(a)$ & Limiting free energy: $\displaystyle \C{F}(a) = \lim_{n\to\infty} \C{F}_n(a)$ \\
$\C{E}_n(a)$, $\C{E}(a)$ & Energy density and limiting energy density \\
$\C{C}_n(a)$, $\C{C}(a)$ & Specific heat and limiting specific heat \\
$H_n$ & The mean height of the endpoint of the walk (a function of $a$) \\
$R_n^2$ & The mean square radius of gyration of the walk (a function of $a$) \\
$P^+(\eps)$ & The microcanonical density function (see equation \Ref{eqn59X}) \\
$a_c^+$ & The adsorption critical point \\
$\phi$ & The crossover exponent \\
$\alpha$ & The specific heat exponent \\
$\gamma_1$ & The half-space entropic exponent, see also $\gamma_s$ (the surface exponent) \\
$\nu$ & The metric exponent \\
$G(a,t)$  & Generating function of $Z_n(a)$ (see equation \Ref{eqn45}) \\
$G_N(a,t)$ & Truncated generating function (see equation \Ref{eqn46a}) \\
$t_c^+(a)$ & The radius of convergence of $G(a,t)$ \\
\hline\hline
\end{tabular}
}
\end{center}
\end{table}
}

\subsection{Organisation of the manuscript}

This paper is a report on two aspects of the Monte Carlo simulation of
adsorbing walks.  The first is the generalisation of the GAS algorithm to 
a model of interacting walks, and in particular, an implementation of this
algorithm to achieve flat histogram sampling over state space of
adsorbing square and cubic lattice walks.  The second aspect of the paper
is a report on the properties of adsorbing self-avoiding walks in the 
square and cubic lattices.  The aim here is to verify the results obtained
in the Multiple Markov Chain Monte Carlo study in reference \cite{JvRR04},
and also to use the data generated here to test the scaling of the
thermodynamic and metric quantities of adsorbing walks.

The model of adsorbing walks is defined in section \ref{subsection1.2}, and
its partition function and free energy are discussed. The limiting free energy of
this model exists, and its properties have been examined
elsewhere \cite{HTW82}; see, for example, reference \cite{JvR15}.  
The basic scaling relations for the free energy, energy, and
specific heat were introduced above, and the scaling of the partition function
was briefly reviewed.

In section \ref{section2} the GAS algorithm \cite{JvRR09} is reviewed and 
then generalised to interacting models.  The algorithm normally has only one 
set of parameters (associated with the size of the walks), but it is shown
here that introducing a second set of parameters (associated with the
energy of the walks) can give an algorithm which samples effectively in
both length and energy.  It is shown that the algorithm can be tuned
to give flat histogram sampling in the spirit of the PERM algorithm 
\cite{G97,HG11} (but without using enrichment or pruning of states).

Numerical results for adsorbing walks are analysed in section \ref{section3}.
The location of the adsorption critical point is determined from the 
mean energy of adsorbing walks, giving
\begin{equation}
a_c^+ =
\cases{
1.779 \pm 0.003, & \hbox{in the square lattice}; \\
1.306 \pm 0.007, & \hbox{in the cubic lattice}.
}
\end{equation} 
These results are then used to estimate the crossover exponent $\phi$
associated with the adsorption transition,  giving
\begin{equation}
\phi =
\cases{
0.496 \pm 0.009, & \hbox{in two dimensions}; \\
0.505 \pm 0.006, & \hbox{in three dimensions}.
}
\end{equation} 
The microcanonical density function of adsorbing walks is determined
as well, and shown to have properties consistent with the location
of the critical points above.  In addition, the specific heat of the 
model is determined, and it is found that it has scaling behaviour consistent
with the estimates of the critical points above.

It is also shown that the adsorption transition is seen in a change in the
metric scaling of walks at the critical point.  The desorbed phase is a 
phase of positive walks with the scaling properties of self-avoiding walks
in a good solvent, while the adsorbed walk has the metric properties of
a walk in one dimension less; that is, of walks adhering to the adsorbing
surface.  

Scaling of the generating and partition functions are found to be 
consistent with the exact values of critical exponents determined elsewhere 
in two dimensions \cite{BEG89,BY95}, and with the value $\phi=\shalf$ 
in three dimensions \cite{HG94,JvRR04,ML88A}.  Similary, the data are
also consistent with estimates for the surface exponent $\gamma_s$; 
this is seen particularly in the scaling of the partition function
(see equation \Ref{eqn15}). 

The paper is concluded with a few brief remarks and a summary in
section \ref{section4}.

\begin{figure}[t]
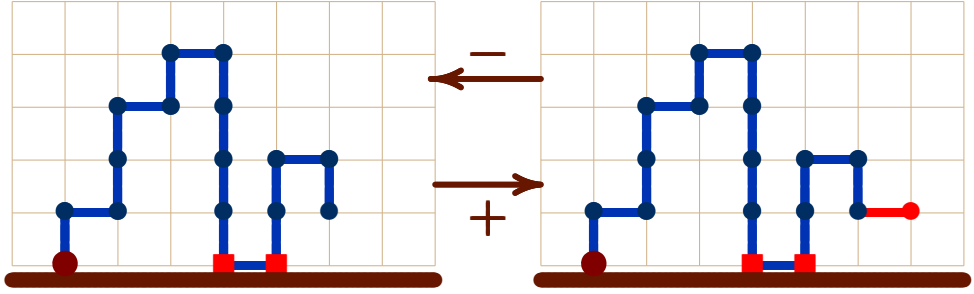

\input figure2-.tex
\caption{Endpoint elementary moves.  Appending an edge at the endpoint of the
walk on the left is a \textit{positive elementary move}.  Removing the last edge
on the right is the reverse of a positive elementary move.  This is a \textit{negative
elementary move}.}
\label{figure1}  %%ZXZ[figure1]
\end{figure}

\section{GAS Sampling of self-avoiding walks}
\label{section2}

The GAS algorithm is a generalisation of kinetic growth algorithms.  It is designed to
sample along weighted sequences in state space in such a way that the ratios
of average weights of sequences ending in walks of length $n$ and $m$ are estimates
of the ratio of the numbers of walks of lengths $n$ and $m$.  In this section I show how
to generalise this algorithm so that it can be used to estimate the number of 
walks of length $n$ and \textit{energy} $m$.  That is, the algorithm will be used to
sample walks in the microcanonical ensemble.

Let $w=\LA \omega_0,\omega_1,\ldots,\omega_n\RA$ be a self-avoiding walk of 
length $n$ from its source vertex $\omega_0 =\vec{0}$ at the origin, to its terminal 
vertex $\omega_n$, giving $n$ steps $\LA \omega_{j-1},\omega_j\RA$ for 
$j=1,2,\ldots,n$.  The walk $w$ may be made longer by adding a step 
$\LA \omega_n,\omega_{n+1}\RA$ to $\omega_n$, or it may be made
shorter by removing its last step.  These two operations compose an 
end-point elementary move for sampling self-avoiding walks, as illustrated
in figure \ref{figure1}:  A \textit{positive elementary move} is the addition of an edge 
to the endpoint of a growing walk.  The reverse of a positive elementary move
is a \textit{negative elementary move}, namely the deletion of the least edge
in a walk.  Notice that every positive move is immediately reversible by a negative
elementary move.

Endpoint elementary moves have been used widely in the simulation of 
self-avoiding walks (for example the Rosenbluth algorithm \cite{RR55}, and 
the Beretti-Sokal algorithm \cite{BS85}).  In what follows the discussion will 
be restricted to endpoint elementary moves; however, the algorithms 
generalise directly if other elementary moves, such as BFACF elementary 
moves \cite{BF81,AA83}, or generalised atmospheric moves
\cite{RJvR08}, are used instead.

\begin{figure}[t]
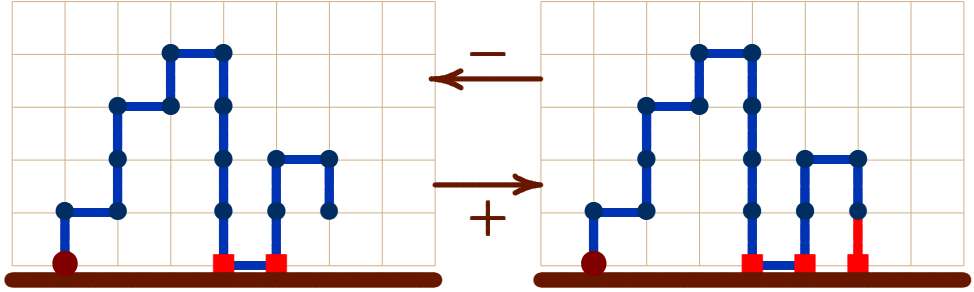

\input figure3-.tex
\caption{A endpoint elementary move may increase or decrease the number of 
visits in a walk.  By appending an edge at the endpoint of the
walk on the left a new visit is created.  Removing the last edge
on the right is the reverse of this move, and it decreases the number of visits by
one.}
\label{figure2}  %%ZXZ[figure2]
\end{figure}

As an example, consider the elementary move in figure \ref{figure1}, which is an
implementation of an endpoint elementary move on a self-avoiding walk in
the positive half-lattice $\mathL^+_2$ (where $\vec{w}(2)$ is the $y$-coordinate 
of $\vec{w}$).  The set of lattice
edges in $\mathL^+_2$ which may be appended to the walk $w$ to extend it
by one step is the \textit{positive atmosphere} of $w$, and the number of
edges in the positive atmosphere is denoted by $a^+(w)$.  For example,
for the walk on the left in figure \ref{figure1}, $a^+(w)=2$.

Similarly, the set of edges which may be removed from the endpoint of
a walk to decrease its length by one, is the \textit{negative atmosphere} of
the walk.  For endpoint elementary moves, the last step is always the sole negative
atmospheric edge, so that the size of the negative atmosphere for endpoint
elementary moves is always $a^-(w)=1$, if $w$ is not the trivial walk of length
$0$.

An elementary move may change the energy of a walk.  For example,
in a model of adsorbing walks in $\mathL^+_2$, the energy is the number of 
\textit{visits} of the walk to the adsorbing line $\partial\mathL^+_2$ (the boundary
of the half-lattice $\mathL^+_2$).  The positive move in figure \ref{figure1} 
does not change the number of visits, but the move in figure \ref{figure2}
increases the number of visits (and so the energy) by $1$.  The negative move
in figure \ref{figure2} similarly reduces the number of visits by $1$.

\begin{figure}[h]
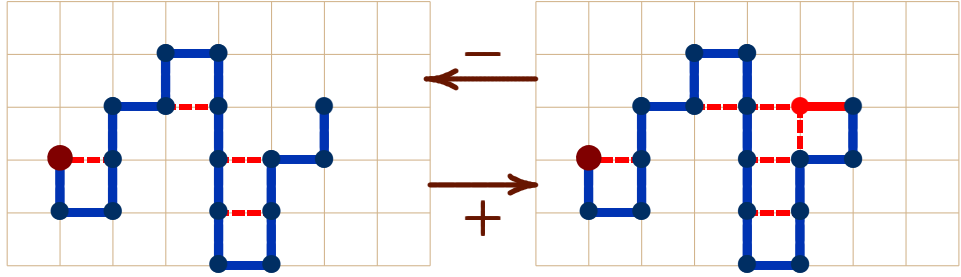

\input figure4-.tex
\caption{A endpoint elementary move may increase or decrease the number of 
 contacts in a walk.  By appending an edge at the endpoint of the
walk on the left two new contacts are created.  Removing the last edge
on the right is the reverse of this move, and it decreases the number of contacts
by two.}
\label{figure3}  %%ZXZ[figure3]
\end{figure}

A similar situation arises if a model of collapsing walks with energy given by
nearest neighbour \textit{contacts} between vertices in the walk which are
adjacent in $\mathL$, but not in the walk.  This is illustrated in figure
\ref{figure3}; the positive elementary move creates $2$ new contacts
in the walk, and so it changes the energy of the walk by $2$.

More general elementary moves (for example, the BFACF elementary moves)
may contribute to the atmospheric
statistics $a^+$ and $a^-$ in various ways, and may even give rise to
\textit{neutral atmospheres} which do not change the length of the walk
(but which may change the energy of the walk).

Thus, in what follows, let $a^+_v(w)$ be the size of the positive 
atmosphere of a walk $w$ of length $\ell_n=|w|$, and which changes the
energy of $w$ by $v$ units.  For example, the walk on the left in figure 
\ref{figure1} has $a^+_{0}(w)=1$ and $a^+_{1}(w)=1$, and if this
was a walk in the three dimensional half-lattice $\mathL_3^+$, then
$a^+_{0}(w)=3$ and $a^+_{1}(w)=1$.  The walk on the left in 
figure \ref{figure3} (with energy given by the number of contacts), 
has, in a similar way, $a^+_{0} (w)=2$,  $a^+_{1}(w)=0$ 
and $a^+_{2} (w)=1$. 

In exactly the same way one may define the size of the negative atmosphere
of a walk $w$ of length $\ell$ which changes the energy by $v$, denoted 
by $a^-_{v} (w)$.
  
The neutral atmosphere (an elementary move which does not change the 
length of the walk) of a walk $w$ is similarly given by $a^0_{v} (w)$, if 
it changes the energy of a walk by $v$.  If the endpoint elementary move in 
figure \ref{figure1} is used, then $a^0_{v} (w)=0$ by default (since there
are no neutral elementary moves implemented), but in general assume that a 
more general set of elementary moves (for example BFACF elementary moves) 
is used to sample walks, and in that case the neutral atmosphere may have 
positive size.

\subsection{Implementation of GAS-sampling}

Suppose that an elementary move is implementated on the state space of
self-avoiding walks from the origin, and assume the implementation is
irreducible (that is, the elementary move gives a connected graph
on the state space of walks).

Suppose that the sequence
\begin{equation}
\phi_N = \LA w_0,w_1,w_2,\ldots,w_n,\ldots,w_N\RA
\label{eqn1}     %%ZXZ[eqn1]
\end{equation}
is realised after $N$ steps and that the atmospheres of the states $w_n$ have
sizes (or \textit{statistics}) $a_{v}^+(w_n)$, $a_{v}^0(w_n)$ and 
$a_{v}^-(w_n)$.   These elementary moves may be classified as follows,
depending on whether they increase or decrease the lengths of the walks
(or are neutral), or whether they increase or decrease, or leave unchanged,
the energy of the walk.  This is done by defining atmospheric statistics as
follows:  Let the states $w_n$ be walks of length $\ell_n = |w_n|$.  Define 
\begin{eqnarray*}
\alpha_n^{--} = \sum_{v< 0} a_{v}^- (w_n),\qq & 
\alpha_n^{-0} = \sum_{v=0} a_{v}^-(w_n), \qq 
\alpha_n^{-+} = \sum_{v>0} a_{v}^-(w_n); \\
\alpha_n^{0-} = \sum_{v< 0} a_{v}^0 (w_n),\qq & 
\alpha_n^{00} = \sum_{v=0} a_{v}^0(w_n), \qq 
\alpha_n^{0+} = \sum_{v>0} a_{v}^0(w_n); \\
\alpha_n^{+-} = \sum_{v< 0} a_{v}^+ (w_n),\qq & 
\alpha_n^{+0} = \sum_{v=0} a_{v}^+(w_n), \qq 
\alpha_n^{++} = \sum_{v>0} a_{v}^+(w_n).  
\end{eqnarray*}
For example,  $\alpha_n^{--}$ is the number of negative elementary moves which
also \textit{decreases} the energy of the walk, and $\alpha_n^{-+}$ 
is the number of negative elementary moves which also \textit{increases}
the energy of the walk.  The rest of the $\alpha$'s are similarly
defined.  

With these atmospheric statistics defined, a rule needs to be constructed in order
to realise the sequence $\phi_N$ in equation \Ref{eqn1}.  

The endpoint elementary moves in figures \ref{figure2} and \ref{figure3}
have the property that no positive elementary move can decrease the
energy, and no negative elementary move can increase the energy.   Moreover,
there are no neutral moves amongst the elementary moves. Thus,
assume that
\begin{equation}
\alpha_n^{-+} = 0,\q\alpha_n^{+-}=0,\q\alpha_n^{0-}=0,\q\alpha_n^{00}=0 
\q\hbox{and}\; \alpha_n^{0+} = 0.
\end{equation}
The algorithm can be modified appropriately to account for such transitions
in models where this is not the case.  This assumption leaves the following 
atmospheric statistics:  $\{\alpha_n^{--},\alpha_n^{-0},\alpha_n^{+0},\alpha_n^{++}\}$.  
That is, distinguish between negative moves which decrease the energy, 
or negative moves which leave the energy unchanged, or positive moves which leave 
the energy unchanged, and positive moves which increase the energy.

Introduce parameters $\{\beta_{\ell,u}\}$  to control positive elementary 
moves on walks of length $\ell$ and energy $u$, and which leave the energy 
unchanged (that is, the elementary moves contributing to
$\alpha_n^{+0}$).   Similarly, introduce the parameters $\{\gamma_{\ell,u}\}$ to
control positive elementary moves which increase the energy
on walks of length $\ell$ and energy $u$.  

The transition 
probabilities of positive elementary moves which leave the energy unchanged
will be proportional to $\beta_{\ell,u}$; if the state has length $\ell$ and
energy $u$.  For example, since $\ell=16$ and $u=2$ in the walk in figure
\ref{figure1}, the transition probability of the positive move in that figure
is proporsional to $\beta_{16,2}$.  Similarly, the positive elementary move
in figure \ref{figure2} increases the energy; so here the transition probability is
proportional to $\gamma_{16,2}$, instead.  In figure \ref{figure3} the 
positive elementary move also increases the energy, and so its transition 
probability is proportional to $\gamma_{16,4}$.

Thus, if $w_n$ is the current state (of length $\ell_n$ and energy $u_n$), 
and $w_{n+1}$ is the next state (of length $\ell_{n+1}$ and energy $u_{n+1}$), then define
the change in length by $\Delta_n = \ell_{n+1}\minus \ell_n$, and the change in
energy by $\delta_n = u_{n+1} \minus u_n$.   Notice that $\Delta_n=\pm 1$ for
endpoint elementary moves, and that $\delta_n = \pm 1$, or $\delta_n = 0$,
for the model of adsorbing walks in figures \ref{figure1} and \ref{figure2} (but 
these quantities may take on other values in the model in figure \ref{figure3}).  The transition
probabilities are chosen such that
\begin{equation}
\Pr(w_n\to w_{n+1}) \propto
\cases{
\beta_{\ell,u}, & \hbox{if $\Delta_n = +1$ and $\delta_n=0$;} \\
1 & \hbox{if $\Delta_n = -1$ and $\delta_n=0$},
}
\end{equation}
where $\ell=\ell_n$ and $u=u_n$ are functions of $n$.
In the case that the energy is changed, then the transition probability is
constructed such that
\begin{equation}
\Pr(w_n\to w_{n+1}) \propto
\cases{
\gamma_{\ell,u}, & \hbox{if $\Delta_n=+1$ and $\delta_n>0$;} \\
1 & \hbox{if $\Delta_n=-1$ and $\delta_n < 0$}.
}
\end{equation}
Normalising the transition probabilities  gives
\begin{equation}
\fl
\Pr(w_n \to w_{n+1})
= \cases{
\frac{\beta_{\ell,u}}{\alpha_n^{--} + \alpha_n^{-0} + \alpha_n^{+0}\thin\beta_{\ell,u} 
+ \alpha_n^{++}\thin\gamma_{\ell,u}}, & \hbox{if $\Delta_n=1$ and $\delta_n=0$}; \\
\frac{\gamma_{\ell,u}}{\alpha_n^{--} +\alpha_n^{-0} + \alpha_n^{+0}\thin\beta_{\ell,u} 
+ \alpha_n^{++}\thin\gamma_{\ell,u}}, & \hbox{if $\Delta_n=1$ and $\delta_n>0$}; \\
\frac{1}{\alpha_n^{--} +\alpha_n^{-0} + \alpha_n^{+0}\thin\beta_{\ell,u} 
+ \alpha_n^{++}\thin\gamma_{\ell,u}}, & \hbox{if $\Delta_n=-1$ and $\delta_n\leq 0$}; \\
}
\label{eqn4}   %%ZXZ[eqn4]
\end{equation}
The probability for any type of move can 
be explicitly computed for any given state $w_n$ (of length $\ell$ and energy $u$)
by computing the $\alpha$'s.
For example, the probability for a positive elementary move which leaves
the energy unchanged is $\alpha_n^{+0}\thin\beta_{\ell,u}/(\alpha_n^{--} \plus \alpha_n^{+0}\thin\beta_{\ell,u} \plus  \alpha_n^{++}\thin\gamma_{\ell,u})$.  For the
state on the left in figure \ref{figure1} this becomes
$\beta_{16,2}/(1+\beta_{16,2}+\gamma_{16,2})$ since $\alpha_{16}^{--}=1$,
$\alpha_{16}^{+0} = 1$ and $\alpha_{16}^{++}=1$.

\subsection{GAS-weights}

Let $w_0$ be a starting state (possibly the walk consisting of a single vertex
at the origin, of length $\ell_0=0$ and energy $u_0=0$).  Implement
the endpoint elementary moves on $w_0$ by computing its atmosphere
recursively and updating it using the transition probabilities in equation \Ref{eqn4},
starting at $n=0$.  This generates a Markov Chain of states in 
a sequence $\phi_N$ (see equation \Ref{eqn1}).

Assume that the parameters $\{\beta_{n,u},\gamma_{n,u}\}$ are known
and fixed, so that the sampling can be implemented by simply computing 
the transtition probabilities and selecting positive and negative elementary
moves with appropriate probabilities. 

The probability of the sequence $\phi_N$ is given by
\begin{equation}
\fl
\Pr (\phi_N) = \prod_{n=0}^{N-1}
\frac{1}{\alpha_n^{--} + \alpha_n^{-0} 
+ \alpha_n^{+0}\thin\beta_{\ell,u} 
+ \alpha_n^{++}\thin \gamma_{\ell,u}}
\; {\prod}_m^\prime \beta_{\ell,u}
\; {\prod}_k^{\prime\prime} \gamma_{\ell,u}
\label{eqn6}            %%ZXZ[eqn6]
\end{equation}
where the primed product ${\prod}^\prime$ is over all the $\beta_{\ell,u}$
for transitions through positive elementary moves leaving the energy unchanged,
and the double primed product ${\prod}^{\prime\prime}$ is over all the
$\gamma_{\ell,u}$ where the transition is a positive elementary move increasing
the energy.  In this expression the $\ell$ and $u$ are functions of
$n$, $m$, and $k$ in each of the products.

A weight $W(\phi_N)$ will be assigned to the sequence $\phi_N$.  In order
to compute the weight, define
\begin{equation}
\fl
\sigma(j,j\plus 1) = \sigma(w_j\to w_{j+1}) = 
\cases{
-1, &\hbox{if $\Delta_j=+1$ and $\delta_j=0$}; \\
+1, &\hbox{if $\Delta_j=-1$ and $\delta_j=0$}; \\
\hspace{0.8em} 0, &\hbox{otherwise}. \\
}
\end{equation}
Similarly, define
\begin{equation}
\fl
\rho (j,j\plus 1) = \rho (w_j\to w_{j+1}) = 
\cases{
-1 , &\hbox{if $\Delta_j=+1$ and $\delta_j>0$}; \\
+1, &\hbox{if $\Delta_j=-1$ and $\delta_j<0$}; \\
\hspace{0.8em} 0, &\hbox{otherwise}. \\
}
\end{equation}
That is, the function $\sigma(j,j\plus 1)$ tracks the negative and positive 
moves along $\phi_N$ where the energy is not changed, and
the function $\rho(j,j\plus 1)$ tracks the negative and positive
transitions along $\phi_N$ where the energy is also changed.

Assign the weight
\begin{equation}
\fl
W(\phi_N) = \LB \Sfrac{\alpha^{--}_0 + \alpha^{-0}_0 + \alpha^{+0}_0 \beta_0 
+ \alpha^{++}_0 \gamma_0}{
\alpha^{--}_N +\alpha^{-0}_N +\alpha^{+0}_N \beta_N+ \alpha^{++}_N \gamma_N} \RB
\prod_{j=0}^{N-1} \beta_{\ell,u}^{\sigma(j,j{+}1)}
\prod_{j=0}^{N-1} \gamma_{\ell,u}^{\rho(j,j{+}1)}
\label{eqn9}            %%ZXZ[eqn9]
\end{equation}
to the sequence $\phi_N$.

The expected value of the weight over sequences of length $N$ from state $w_0$ to state
$w_N$ is
\begin{equation}
\LA W(w_0\to w_N) \RA_N = \sum_{\phi:w_0 \to w_N}  \Pr(\phi) W(\phi) .
\label{eqn10}    %%ZXZ[eqn10]
\end{equation}
Inserting equations \Ref{eqn6} and \Ref{eqn9} in this, and simplifying, gives
\begin{equation}
\fl
\LA W(w_0\to w_N) \RA_N = \sum_{\phi:w_0\to w_N}
\prod_{j=1}^N 
\frac{1}{\alpha_j^{--} \plus \alpha_j^{-0} \plus \alpha_j^{+0}\thin\beta_{\ell,u} 
\plus  \alpha_j^{++}\thin \gamma_{\ell,u}}
\; {\prod}^{n} \beta_{\ell,u}
\; {\prod}^{nn} \gamma_{\ell,u}
\label{eqn11}    %%ZXZ[eqn11]
\end{equation}
where the product ${\prod}^{n}$ is over all the $\beta_{\ell,u}$
for transitions through \textit{negative} elementary moves leaving the energy unchanged,
and the product ${\prod}^{nn}$ is over all $\gamma_{\ell,u}$ where the transition is a 
\textit{negative} elementary move decreasing the energy.  As before, the $\ell
\equiv \ell(j)$ and $u\equiv u(j)$ are functions of $j$ (in other words, 
functions of the states $w_j$ in the sequence $\phi$).

Reverse all the sequences in equation \Ref{eqn11} so that the starting state is $w_N$ and
the final state is $w_0$.  Under this reversal all negative elementary moves become 
positive elementary moves and vice versa.  That is, equation \Ref{eqn11} becomes
\begin{equation}
\fl
\LA W(w_0\to w_N) \RA_N = \sum_{\psi:w_N\to w_0}
\prod_{j=1}^N 
\frac{1}{\alpha_j^{--} \plus \alpha_j^{-0} \plus \alpha_j^{+0}\thin\beta_{\ell,u} 
\plus \alpha_j^{++}\thin \gamma_{\ell,u}}
\; {\prod}^{n} \beta_{\ell,u}
\; {\prod}^{nn} \gamma_{\ell,u} ,
\label{eqn12}    %%ZXZ[eqn12]
\end{equation}
where, as before, the product ${\prod}^{n}$ is over all the $\beta_{\ell,u}$
for transitions through \textit{negative} elementary moves along the
\textit{reverse sequence} $\psi$ leaving the energy unchanged,
and the product ${\prod}^{nn}$ is over all $\gamma_{\ell,u}$ where the transition is a 
\textit{negative} elementary move along the \textit{reverse sequence} $\psi$
decreasing the energy.

For example, consider the model of collapsing walks in figure \ref{figure3} 
and suppose the sequence $\phi$ is realised, where

\beginpicture
\put {$\phi = \hbox{\LARGE $\langle$}$} at -25 5
\put {\Large .} at 0 0
\put {$\rightarrow$} at 20 0 
\plot 40 0 50 0 / \multiput {\Large .} at 40 0 50 0 /
\put {$\rightarrow$} at 70 0 
\plot 90 0 100 0 100 10 / \multiput {\Large .} at 90 0 100 0 100 10 /
\put {$\rightarrow$} at 120 0 
\plot 140 0 150 0 150 10 140 10 / \multiput {\Large .} at 140 0 150 0 150 10 140 10 /
\put {$\rightarrow$} at 170 0 
\plot 190 0 200 0 200 10 / \multiput {\Large .} at 190 0 200 0 200 10 /
\put {$\rightarrow$} at 220 0 
\plot 240 0 250 0 250 10 240 10 / \multiput {\Large .} at 240 0 250 0 250 10 240 10 /
\put {$\rightarrow$} at 270 0 
\plot 300 0 310 0 310 10 300 10 290 10 / \multiput {\Large .} at 300 0 310 0 310 10 300 10 290 10 /
\put {\hbox{\LARGE $\rangle$.}} at 320 5 
\put {$w_0$} at 0 -10
\put {$w_1$} at 45 -10
\put {$w_2$} at 95 -10
\put {$w_3$} at 145 -10
\put {$w_4$} at 195 -10
\put {$w_5$} at 245 -10
\put {$w_6$} at 300 -10
\put { } at 0 -20
\endpicture

The probability of this sequence is
\begin{equation}
\fl
\Pr(\phi) = \Sfrac{\beta_{0,0}}{4\beta_{0,0}}
\Sfrac{\beta_{1,0}}{(1+3\thin\beta_{1,0})}
\Sfrac{\gamma_{2,0}}{(1+2\thin\beta_{2,0}+\gamma_{2,0})}
\Sfrac{1}{(1+2\thin\beta_{3,1})}
\Sfrac{\gamma_{2,0}}{(1+2\thin\beta_{2,0}+\gamma_{2,0})}
\Sfrac{\beta_{3,1}}{(1+2\thin\beta_{3,1})}
\end{equation}

The weight of $\phi$ can similarly be computed from equation \Ref{eqn9}.
This gives
\begin{equation}
W(\phi) = \Sfrac{4\thin\beta_{0,0}}{1+3\thin\beta_{4,1}}
\times \LB  \beta_{0,0}^{-1}\,\beta_{1,0}^{-1}\,\gamma_{2,0}^{-1}
\,\gamma_{3,1}^{+1}\,\gamma_{2,0}^{-1}\,\beta_{3,1}^{-1}\RB .
\end{equation}
The consequence is that
\begin{equation}
\fl
\Pr(\phi)W(\phi) = 
\Sfrac{1}{(1+3\thin\beta_{4,1})}
\Sfrac{1}{(1+2\thin\beta_{3,1})}
\Sfrac{1}{(1+2\thin\beta_{2,0}+\gamma_{2,0})}
\Sfrac{\gamma_{3,1}}{(1+2\thin\beta_{3,1})}
\Sfrac{1}{(1+2\thin\beta_{2,0}+\gamma_{2,0})}
\Sfrac{1}{(1+3\thin\beta_{1,0})} ,
\end{equation}
and this is the probability that a sequence $\psi$, starting in the state $w_6$ and terminating
in the state $w_0$, is realised by the algorithm.

The same observation is true generally for equation \Ref{eqn12}:  The summand 
is the probability that a particular sequence $\psi$ from state $w_N$ to $w_0$ is
realised by the algorithm, and the summation is over all such sequences $\psi$.
That is, $\LA W(w_0\to w_N) \RA_N = \Pr(w_N\to w_0)$ is the probability that the algorithm 
realises a sequence $\psi$ of length $N$ from state $w_N$ to state $w_0$.  

The sequence $\psi$ is a Markov Chain, and if it is aperiodic and irreducible, and
if $w_0$ is a recurrent state, then asymptotically (for large $N$) the probability
that the sequence hits the state $w_0$ is positive and independent of the starting state
$w_N$.  That is, $\Pr(w_N\to w_0) \to C(w_0)>0$ as $N\to\infty$ where $C(w_0)$ is dependent
on the parameters of the algorithm, and independent of $w_N$.   Thus, the
average weight $\LA W(w_0\to w_N) \RA_N \to C(w_0)$ as $N\to\infty$.  Summing
$\LA W(w_0\to w_N) \RA_N$ over all the states $w_N$ of length $n = \ell(w_N)$
and energy $u$, shows that the average weights of sequences of length $N$ 
ending in walks of length $n$ and energy $u$ is
\begin{equation}
\LA W_{n,u} \RA_N = \sum_{w_N:\,n,u}  \LA W(w_0\to w_N) \RA_N \to c_n(u)\, C(w_0) ,
\end{equation}
where the summation is over all walks of length $n$ and energy $u$ (and
$c_n(u)$ is the number of walks of length $n$ and energy $u$).

Taking ratios of average weights give
\begin{equation}
\frac{\LA W_{n,u} \RA_N}{\LA W_{m,v} \RA_N} \to \frac{c_n(u)}{c_m(v)} .
\end{equation}
That is, if $c_n(u)$ is known for some values of $n$ and $u$, then the ratios
of average weights can be used to estimate $c_m(v)$.

{\renewcommand{\baselinestretch}{1.5}
% Numerical estimates of $\beta_{n,u}$
\begin{table}[t!]
\begin{center}
{
\caption{Numerical estimates of $\beta_{n,u}$ for adsorbing walks in $\mathL^2_+$}
\label{betanu}  %%ZXZ[betanu]
}
{
\begin{tabular}{l|lllllllllll}
\hline\hline
$n\backslash u$ & 0 & 1 & 2 & 3 & 4 & 5 & 6 &7 & 8 & 9 & 10 \\
\hline
   0 & \fns 1.000 \\
   1 & \fns 0.333  & \fns 1.000 \\
   2 & \fns 0.428 & \fns 0.250  & \fns 1.001 \\
   3 & \fns 0.368 & \fns 0.499 & \fns 0.200  & \fns 1.000 \\
   4 & \fns 0.387 & \fns 0.380 & \fns 0.416 & \fns 0.166  & \fns 1.000 \\
   5 & \fns 0.373 & \fns 0.396 & \fns 0.427 & \fns 0.429 & \fns 0.142  & \fns 1.000 \\ 
   6 & \fns 0.386 & \fns 0.375 & \fns 0.377 & \fns 0.359 & \fns 0.438 & \fns 0.124  & \fns 1.000 \\
   7 & \fns 0.377 & \fns 0.392 & \fns 0.384 & \fns 0.401 & \fns 0.348 & \fns 0.445 & \fns 0.111  & \fns 1.000 \\
   8 & \fns 0.383 & \fns 0.378 & \fns 0.388 & \fns 0.356 & \fns 0.368 & \fns 0.327 & \fns 0.449 & \fns 0.100 & \fns 0.999 \\
   9 & \fns 0.378 & \fns 0.386 & \fns 0.385 & \fns 0.395 & \fns 0.370 & \fns 0.368 & \fns 0.308 & \fns 0.454 & \fns 0.0908 & \fns 0.999 \\
  10 & \fns 0.381 & \fns 0.379 & \fns 0.383 & \fns 0.373 & \fns 0.368 & \fns 0.352 & \fns 0.364 & \fns 0.289 & \fns 0.458 & \fns 0.083 & \fns 0.999 \\
\hline\hline
\end{tabular}
}
\end{center}
\end{table}
}

\subsection{Sampling with GAS}

The algorithm is implemented by choosing a starting state $w_0$ and then
sampling along a sequence $\phi$ of length $N$.  The weight is updated along 
$\phi$, and collected into bins for walks of length $n$ and energy u.  Once the
sequence is completed, then the average of each bin is computed, giving the
average weight $W_{n,u}$ of walks of length $n$ and energy $u$ seen along
$\phi$.

This process is repeated $M$ times, so that $M$ sequences of length $N$,
denoted by $\LA \phi_1, \phi_2,\ldots,\phi_M\RA$ are realised, and for each
sequence $\phi_j$ the average weight $W_{n,u}^{(j)}$ is calculated.  The estimated
average weight is computed over the $M$ sequences:
\begin{equation}
[W_{n,u}]^{est}_{N,M} = \Sfrac{1}{M} \sum_{j=1}^M W_{n,u}^{(j)} .
\end{equation}
The estimated weight $[W_{n,u}]_{N,M}^{est}$ is an estimator of $\LA W_{n,u}\RA_N$,
and so as $M\to\infty$ and $N\to\infty$, it is expected that $[W_{n,u}]_{N,M}^{est}
\to c_n(u)\,C(w_0)$.  Taking ratios give
\begin{equation}
\frac{[W_{n,u}]_{N,M}^{est}}{[W_{m,v}]_{N,M}^{est}} \to \frac{c_n(u)}{c_m(v)} .
\end{equation}

The sampling requires that both $N$ and $M$ are sufficiently large, and there is a
trade-off between these quantities.  $M$ should be large enough to have sufficient
independent estimates of the weights to compute sound averages, and $N$ 
should be large enough to have sufficiently long sequences to have sampled
large enough regions of state space.

{\renewcommand{\baselinestretch}{1.5}
% Numerical estimates of $\gamma_{n,u}$
\begin{table}[t!]
\begin{center}
{
\caption{Numerical estimates of $\gamma_{n,u}$ for adsorbing walks in $\mathL^2_+$}
\label{gammanu}  %%ZXZ[gammanu]
}
{
\begin{tabular}{l|lllllllllll}
\hline\hline
$n\backslash u$ & 0 & 1 & 2 & 3 & 4 & 5 & 6 &7 & 8 & 9 & 10 \\
\hline
   0 & \fns  0. 499 \\
   1 & \fns  0. 499 & \fns  1. 000 \\
   2 & \fns  0. 374 & \fns  1. 001 & \fns  0. 999 \\
   3 & \fns  0. 436 & \fns  0. 801 & \fns  0. 998 & \fns  0. 999 \\
   4 & \fns  0. 451 & \fns  0. 668 & \fns  0. 832 & \fns  0. 998 & \fns  1. 000 \\
   5 & \fns  0. 461 & \fns  0. 751 & \fns  0. 857 & \fns  0. 855 & \fns  1. 000 & \fns  0. 999 \\
   6 & \fns  0. 463 & \fns  0. 716 & \fns  0. 721 & \fns  0. 873 & \fns  0. 874 & \fns  0. 999 & \fns  0. 999 \\
   7 & \fns  0. 470 & \fns  0. 734 & \fns  0. 767 & \fns  0. 845 & \fns  0. 887 & \fns  0. 889 & \fns  0. 998 & \fns  1. 000 \\
   8 & \fns  0. 470 & \fns  0. 728 & \fns  0. 709 & \fns  0. 775 & \fns  0. 834 & \fns  0. 898 & \fns  0. 899 & \fns  0. 999 & \fns  1. 00 \\
   9 & \fns  0. 473 & \fns  0. 741 & \fns  0. 721 & \fns  0. 807 & \fns  0. 835 & \fns  0. 845 & \fns  0. 908 & \fns  0. 907 & \fns  1. 000 & \fns  0. 999 \\
  10 & \fns  0. 474 & \fns  0. 736 & \fns  0. 699 & \fns  0. 752 & \fns  0. 795 & \fns  0. 836 & \fns  0. 853 & \fns  0. 915 & \fns  0. 915 & \fns  1. 000 & \fns  0. 999 \\
\hline\hline
\end{tabular}
}
\end{center}
\end{table}
}

There remains the additional issue of the GAS parameters $\beta_{n,u}$ and
$\gamma_{n,u}$.  These can be estimated using training runs prior to the
simulation. Best results are obtained when the sampling is \textit{flat}, so that
states of size and energy $\{n,u\}$ are sampled uniformly in $\{n,u\}$. This is
best achieved when the GAS sequences are random walks on $n$ and $u$.
Thus, the probability of a positive move should be, on average, equal to the
probability of a negative move.  A good choice for $\beta_{n,u}$ is
\begin{equation}
\beta_{n,u} = \frac{\LA \alpha_{n,u}^{-0}\RA}{\LA \alpha_{n,u}^{+0}\RA }
\label{eqn40C}  %%ZXZ[eqn40C]
\end{equation}
where $\LA\alpha_{n,u}^{-0}\RA$ is the average negative atmosphere which does not
decrease the energy, and $\LA\alpha_{n,u}^{+0}\RA$ is the average positive 
atmosphere which does not increase the energy, of walks of length $n$ and energy $u$.

A similar argument shows that a good choice for $\gamma_{n,u}$ is
\begin{equation}
\gamma_{n,u} = \frac{\LA \alpha_{n,u}^{--}\RA}{\LA \alpha_{n,u}^{++}\RA }
\label{eqn41C}  %%ZXZ[eqn41C]
\end{equation}
where $\LA\alpha_{n,u}^{--}\RA$ is the average negative atmosphere which 
decreases the energy, and $\LA\alpha_{n,u}^{++}\RA$ is the average positive 
atmosphere which increases the energy, of walks of length $n$ and energy $u$.

{\renewcommand{\baselinestretch}{1.5}
% The frequency of states sampled to $n=10$ and $u=10$
\begin{table}[t!]
\begin{center}
{
\caption{The frequency of states sampled to $n=10$ and $u=10$ for adsorbing walks in $\mathL^2_+$}
\label{statessampled}  %%ZXZ[statessampled]
}
{
\begin{tabular}{l|lllllllllll}
\hline\hline
$n\backslash u$ & 0 & 1 & 2 & 3 & 4 & 5 & 6 &7 & 8 & 9 & 10 \\
\hline
0 & \fns 8658 \\
1 & \fns 8616 & \fns 13208 \\
2 & \fns 9642 & \fns 7711 & \fns 13261 \\ 
3 & \fns 8971 & \fns 9622 & \fns 7144 & \fns 13012 \\
4 & \fns 8919 & \fns 9576 & \fns 9261 & \fns 6528 & \fns 12883 \\
5 & \fns 8957 & \fns 8947 & \fns 9501 & \fns 9388 & \fns 5998 & \fns 13072 \\
6 & \fns 8982 & \fns 8736 & \fns 9220 & \fns 9019 & \fns 9281 & \fns 6055 & \fns 13263 \\
7 & \fns 9059 & \fns 8562 & \fns 9026 & \fns 9095 & \fns 8711 & \fns 9733 & \fns 5812 & \fns 13278 \\
8 & \fns 8996 & \fns 8398 & \fns 9030 & \fns 8844 & \fns 8568 & \fns 8914 & \fns 9493 & \fns 5728 & \fns 13362 \\
9 & \fns 9169 & \fns 8162 & \fns 8788 & \fns 8772 & \fns 8579 & \fns 8843 & \fns 8417 & \fns 9689 & \fns 5678 & \fns 13204 \\
10& \fns 9134 &\fns 8182 & \fns 8580 & \fns 8434 & \fns 8559 & \fns 8898 & \fns 8390 & \fns 8221 & \fns 10101 & \fns 5477 & \fns12883 \\
\hline\hline
\end{tabular}
}
\end{center}
\end{table}
}
Computed values for $\beta_{n,u}$ and $\gamma_{n,u}$ for a square lattice adsorbing
walk model in figure \ref{figure1} are given in tables \ref{betanu} and \ref{gammanu}
for $0\leq n\leq 10$ and $0 \leq u \leq 10$.  These data are chopped at an accuracy of
three decimal places.

Observe that $\beta_{n,n} = 1$ (when rounded) and $\gamma_{n,n}=1$.  
This is expected, since there are exactly two states if $n>0$  and $n=u$ 
(a completely adsorbed walk which never leaves the adsorbing line).

With these values of $\beta_{n,u}$ and $\gamma_{n,u}$ the sampling is
reasonably flat, as shown in table \ref{statessampled}.  The data in table
\ref{statessampled} is the number of times a sequence of length $10^6$ visited states
of length $n$ and energy $u$.  The maximum length was set at $n=500$, so that 
the number of pairs $(n,u)$ is $124750$; thus, the expected number of visits to
states of length $n$ and energy $u$ is roughly $8000$.  The data in table
\ref{statessampled} are spread around this number, and the distribution
is a reasonably flat histogram.  There are exceptions for data
along the diagonal where a larger number is seen.  The explanation
for this is that the sequence can only visit states along the diagonal if it starts in 
$(n,u)=(0,0)$ and then stay on the diagonal (doing a random walk on states of
length and energy both equal to $n$) -- the elementary moves chosen in the
simulation do not include neutral moves, and so the sequence cannot enter
the diagonal except at $n=0$.

Observe that since $c_n^+(n)=2$ for $n>0$ in this model, that these states are
rare, but are sampled frequently by the algorithm.  This is an example of
\textit{rare event} sampling, where an algorithm spends significant time sampling
rare states in the tails of a distribution in order to get good estimates of
microcanonical quantities.

Simulations were performed by collecting data over $500$ realised
sequences, each of length $10^9$.  Numerical estimates of $c_{n,u}$ are 
shown in table \ref{countsads2d}.  The data in table
\ref{countsads2d} are rounded to the nearest integer.  These data can be 
compared to exact counts to verify the algorithm and its implementation.

{\renewcommand{\baselinestretch}{1.5}
% Estimates of \mu for walks
\begin{table}[t!]
\begin{center}
{
\caption{Estimates of $c_n^+(u)$ to $n=10$ and $u=10$ for adsorbing walks in $\mathL^2_+$}
\label{countsads2d}  %%ZXZ[countsads2d]
}
{
\begin{tabular}{l|lllllllllll}
\hline\hline
$n\backslash u$ & 0 & 1 & 2 & 3 & 4 & 5 & 6 &7 & 8 & 9 & 10 \\
\hline
  0 & 1 \\
   1 & 1 & 2 \\
   2 & 3 & 2 & 2 \\
   3 & 7 & 8 & 2 & 2 \\
   4 & 19 & 16 & 10 & 2 & 2 \\
   5 & 49 & 42 & 24 & 12 & 2 & 2 \\
   6 & 131 & 106 & 56 & 28 & 14 & 2 & 2 \\
   7 & 339 & 283 & 148 & 76 & 32 & 16 & 2 & 2 \\
   8 & 897 & 720 & 385 & 193 & 92 & 36 & 18 & 2 & 2 \\
   9 & 2338 & 1905 & 990 & 543 & 249 & 110 & 40 & 20 & 2 & 2 \\
  10 & 6178 & 4932 & 2571 & 1372 & 672 & 298 & 130 & 44 & 22 & 2 & 2 \\
\hline\hline
\end{tabular}
}
\end{center}
\end{table}
}

Partition functions (see equation \Ref{eqnZ})
can be directly estimated from the microcanonical 
data in table \ref{countsads2d}, for example, for $n=6$ the partition function
is approximated by
\begin{equation}
Z_6(a) = 131 + 106\thin a+ 56\thin a^2 + 28\thin a^3 + 14\thin a^4 + 2\thin a^5 + 2\thin a^6
\end{equation}
where $a=e^{\upsilon/kT}$ is a Boltzmann factor ($\upsilon$ is the interaction 
energy associated with a single visit).

Additional data were collected in the square and cubic lattices.  For example, 
the average mean square radius of gyration of walks of length $n$ and 
energy $u$ were also determined, as was the average height of the endpoint 
of the walks.  In addition, data were also collected on the average positive and 
negative atmospheric statistics from which estimates of the $\beta_{n,u}$
and $\gamma_{n,u}$ were made (see equations \Ref{eqn40C} and \Ref{eqn41C}).  

\section{Numerical Results}
\label{section3}

\subsection{Adsorbing walks in the square lattice}

\begin{figure}[t]
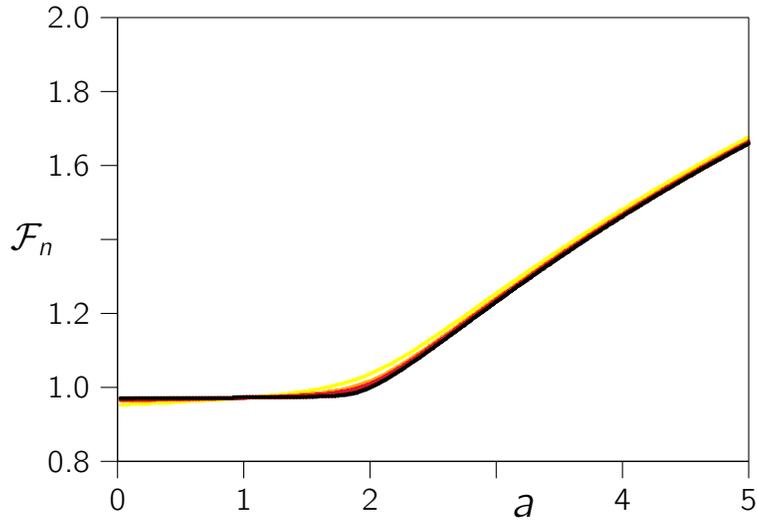

\input figure5-.tex
\caption{The finite size free energy $\C{F}_n(a)$ of adsorbing
positive walks as a function of $a$ for  adsorbing walks in $\mathL_2^+$.  
The lengths of the walks varied
from $n=50$ (yellow) with colours increasing in hue to black when
$n=500$ in steps of $50$. For small values of $a$
$\C{F}_n(a)$ converges to $\log \mu_2$, but for $a$ large
$\C{F}_n(a) > \log \mu_2$.}
\label{figure4}  %%ZXZ[figure4]
\end{figure}

The finite size free energy $\C{F}_n(a)$ was determined from the data
and is plotted in figure \ref{figure4} for walks of lengths 
$n=50\thin N$ with $=1,2,3,\ldots,10$.  $\C{F}_n(a)$ is a function of the 
combined variable $\tau=n^{1/2}(a\minus a_c^+)$
(see equations \Ref{eqn24} and \Ref{eqn26}, and note that the 
crossover exponent is $\phi=\sfrac{1}{2}$).  Plotting the free energy
$\C{F}_n(a)$ against $\tau$ should collapse the data to a single underlying
curve (near the critical point $a_c^+$; that is, for small values of $\tau$) 
which exposes the scaling function $g$ in equation \Ref{eqn26}.  
This is displayed in figure \ref{figure4B}.

\begin{figure}[b]
\input figure6-.tex
\caption{The scaled free energy $n\thin (\C{F}_n - \log \mu_2)$ 
as a function of $\tau = n^{1/2}(a\minus a_c^+)$, where $a_c^+=1.78$,
for  adsorbing walks in $\mathL_2^+$.
The lengths of the walks varied from $n=50$ (yellow) with colours 
increasing in hue to black when $n=500$ in steps of $50$.  The data
collapse for small $|\tau|$ (this is the critical scaling regime), 
and will approach a limiting curve as $n\to\infty$, also for large $\tau$,
given by the limiting free energy. }
\label{figure4B}  %%ZXZ[figure4B]

\end{figure}

\begin{figure}[t]
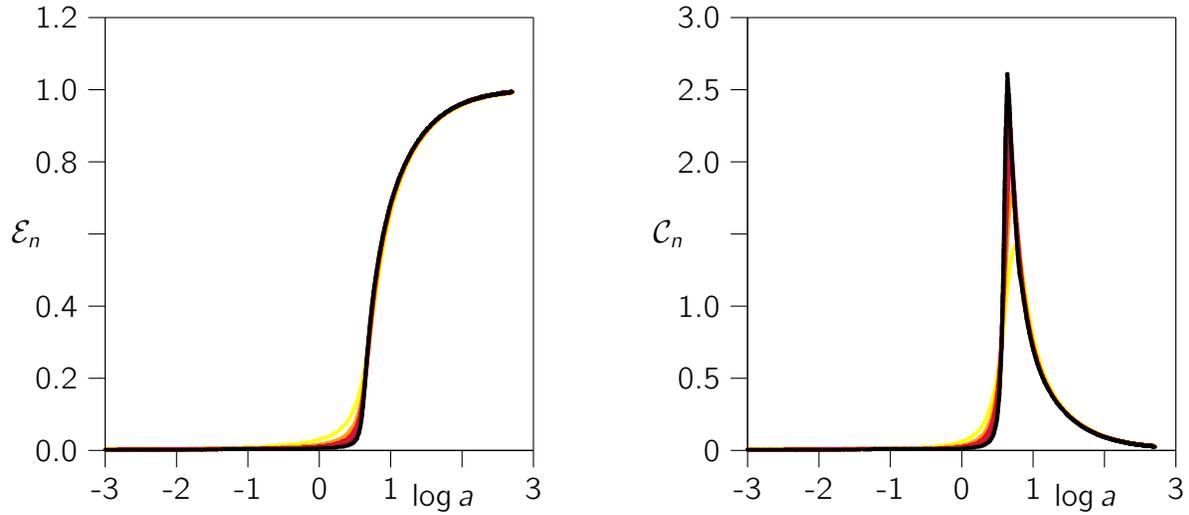

\input figure7-.tex
\caption{The energy $\C{E}_n(a)$ and specific heat $\C{C}_n(a)$
(see equation \Ref{eqn24EC}) as a function of $\log a$ for
adsorbing walks in $\mathL_2^+$.  The lengths of the walks varied
from $n=50$ (yellow) with colours increasing in hue to black when
$n=500$ in steps of $50$. }
\label{figure5}  %%ZXZ[figure5]
\end{figure}

\begin{figure}[b]
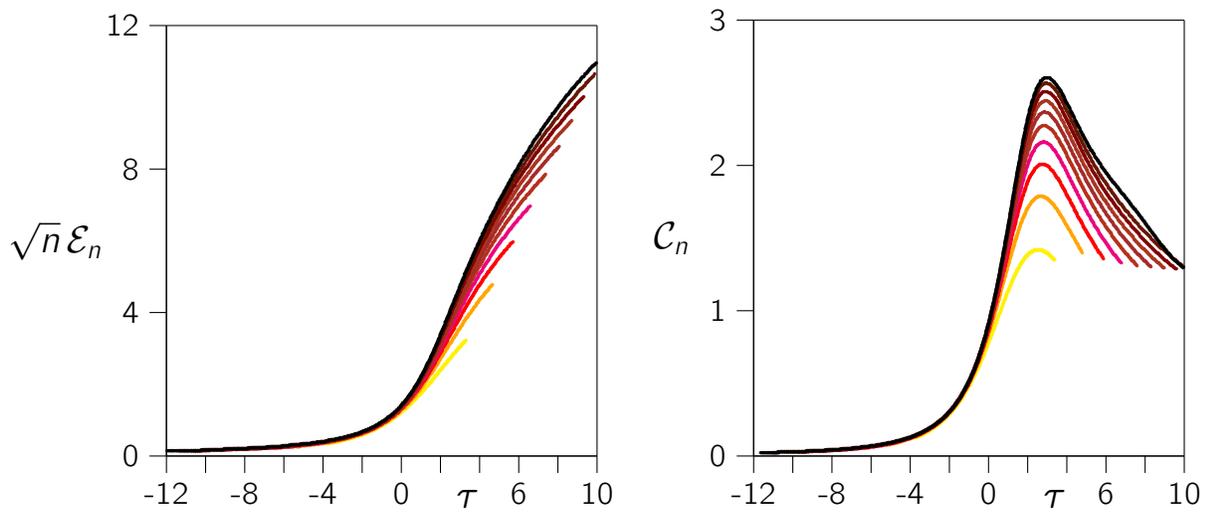

\input figure8-.tex
\caption{The rescaled energy density (left) and the specific heat (right)
plotted as a function of $\tau = n^{1/2}(a\minus a_c^+)$ near the critical
point $a_c^+$ for  adsorbing walks in $\mathL_2^+$.  
The curves approach a limiting curve with increasing $n$. 
The lengths of the walks varied from $n=50$ (yellow) with colours increasing 
in hue to black when $n=500$ in steps of $50$.  The data
collapse for small values of $|\tau|$ (this is the critical scaling regime
which contains the critical point).  For larger values of $|\tau|$ the
curves will approach a limiting curve as $n\to\infty$
given by the limiting free energy.}
\label{figure6}  %%ZXZ[figure6]
\end{figure}

The energy $\C{E}_n(a)$ and specific heat $\C{C}_n(a)$
(see equation \Ref{eqn24EC}) can be determined by differentiation 
of $\C{F}_n(a)$ and is plotted in figure \ref{figure5} against $\log a$
and in figure \ref{figure6} against $\tau$.  These curves clearly show the
adsorption transition in the model at a critical value of $a$.

\begin{figure}[t]
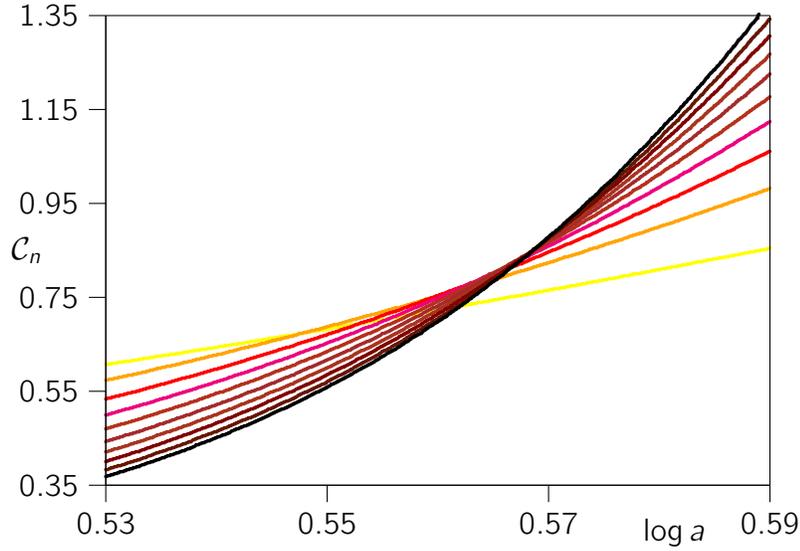

\input figure9-.tex
\caption{A magnification of the point close to where the specific
heat curves in figure \ref{figure5} intersect, 
for  adsorbing walks in $\mathL_2^+$.  The lengths of the walks varied
from $n=50$ (yellow) with colours increasing in hue to black when
$n=500$ in steps of $50$. }
\label{figure5B}  %%ZXZ[figure5B]
\end{figure}

Closer inspection of the specific heat curves in figure \ref{figure5}
shows that they intersect each other close to a fixed point.  To the
left of this point the curves decrease with increasing $n$ to
zero, and to the right of this point the curves increase with increasing
$n$.  The common point of intersection is located approximatedly
at the critical adsorption point $a_c^+$ in the model.  In figure
\ref{figure5B} the specific heat curves are magnified in a region
close to the point where they intersect.

\subsubsection{Location of the critical point $a_c^+$:}
\label{section311}  %%ZXZ[section311]
In order to determine $a_c^+$, consider the finite size energy density 
$\C{E}_n(a) = a\sfrac{d}{da} \C{F}(a)$ per edge in the model.  It is
known that (see, for example, equation \Ref{eqn24EC} and reference \cite{JvRR04})
\begin{equation}
\C{E}_n(a) = n^{\phi-1}\,h_n(\tau)
\end{equation}
where $h_n$ is a scaling function.
From this equation, construct the ratio
\begin{equation}
\frac{\log (n\, \C{E}_n(a))}{\log (m\,\C{E}_m(a))}
= \frac{\phi\log n + \log h_n(\tau)}{\phi\log m + \log h_m(\tau)}
\end{equation}
If $a=a_c^+$, then $\tau = 0$, and the above simplifies to 
\begin{equation}
\frac{\log (n\, \C{E}_n(a_c^+))}{\log (m\,\C{E}_m(a_c^+))}
= \frac{\phi\log n + \log h_n(0)}{\phi\log m + \log h_m(0)} .
\end{equation}
For large values of $n$ and $m$, $\log h_n(0)$ and $\log h_m(0)$
approaches the same constant $C$, so that the above becomes
\begin{equation}
\frac{\log (n\,\C{E}_n(a_c^+))}{\log (m\, \C{E}_m(a_c^+))}
= \frac{\phi\log n + C}{\phi\log m + C} + \hbox{small correction}.
\end{equation}
When $C$ is small compared to $\log m$, the right hand side may be
expanded to obtain
\begin{equation}
\hspace{-1cm}
\frac{\log (n\,\C{E}_n(a_c^+))}{\log (m\, \C{E}_m(a_c^+))}
= \Sfrac{\log n}{\log m} + \LH \Sfrac{C}{\phi \log m}
- \Sfrac{C \log n}{\log^2 m} \RH - 
\Sfrac{C^2}{\phi^2\log^2 m} + \hbox{small correction}.
\end{equation}
The signs of the two terms in square brackets are opposite, and for $n$
and $m$ not too far apart, these terms grow about at the same rate
as the last term $\sfrac{C^2}{\phi^2\log^2 m}$.  That is, the approximation
\begin{equation}
\hspace{-1cm}
\frac{\log (n\,\C{E}_n(a_c^+))}{\log (m\, \C{E}_m(a_c^+))}
= \Sfrac{\log n}{\log m}  - 
\Sfrac{C_0}{\phi^2\log^2 m} + \hbox{small correction},
\end{equation}
where $C_0$ is a constant, should be accurate at the critical point.  
Divide both sides by $\sfrac{\log n}{\log m}$ to obtain
\begin{equation}
\hspace{-1cm}
P_{n,m}(a_c^+) 
= \frac{\log (n\,\C{E}_n(a_c^+))}{\log (m\, \C{E}_m(a_c^+))}\frac{\log m}{\log n}
= 1 - \Sfrac{C^2}{\phi^2\log n \log m} + \hbox{small correction},
\label{eqn44C}   %%ZXZ[eqn44C]
\end{equation}
Solving for the critical point by inverting $P_{n,m}$ gives the solution $a_{n,m}^+$ as
an approximation of the critical point, namely
$a_{n,m}^+ 
= P_{n,m}^{-1} (1 \minus \sfrac{C^2}{\phi^2\log n \log m} \plus \hbox{small correction})$
which  may be expanded to
\begin{equation}
a_{n,m}^+ = P_{n,m}^{-1}(1) - \Sfrac{C_1}{\log n \log m} +
\hbox{small correction},
\label{eqn44D}  %%ZXZ[eqn44D]
\end{equation}
where $C_1$ is a constant.  That is, for given values of $n$ and $m$, an 
estimate of $a_c^+$ can be obtained by determining the solution of
$P_{n,m}(a) = 1$, or $a_{n,m}^+$ is estimated obtained
by solving for $a$ in 
\begin{equation}
\hspace{-1cm}
\frac{\log (n\,\C{E}_n(a))}{\log (m\, \C{E}_m(a))}\frac{\log m}{\log n}
= 1.
\end{equation}
In the above, $m$ was put equal to $n\minus 100$, and $n$ was assigned 
values starting at $n=200$ to $n=500$ in steps of $1$.  The estimates 
$a_n^+ \equiv a_{n,n-100}^+$ showed no systematic dependence on $n$, 
and a simple average over all $n\in[200,500]$ gives the best value
\begin{equation}
a_c^+ = 1.7788 \pm 0.0029 .
\label{abest}   %%ZXZ[abest]
\end{equation}
The confidence interval is obtained by doubling the square root of
the variance of the estimates $a_n^+$.  This result compares 
well with the result in reference \cite{BGJ12}, namely 
$a_c^+ = 1.77564$ (obtained by the exact enumeration of adsorbing walks),
and also with $a_c^+ = 1.759\pm 0.018$ in reference \cite{JvRR04} (obtained by
using a Multiple Markov Chain implementation of the Berretti-Sokal algorithm
\cite{BS85}).

The estimate \Ref{abest} can be used to determine the crossover exponent
$\phi$.  By equation \Ref{eqn24EC}, the specific heat scales as
$\C{C}_n(a_c^+) \sim n^{\alpha\phi} h_c(0)$ when $a=a_c^+$.  That is,
an estimate of $\alpha\phi$ is obtained by computing 
$\sfrac{\log \C{C}_n(a_c^+)}{\log n}$.  Computing this for $100 \leq n \leq 500$,
and taking the mean as the best estimate (and estimating a confidence interval 
by doubling the square root of the variance of the estimates), gives
$\alpha\phi = -0.0091 \pm 0.0162$.  Determining $\phi$ by using equation
\Ref{eqn25} then gives the best estimate for $\phi$:
\begin{equation}
\phi = 0.4955 \pm 0.0081 .
\label{phibest}   %%ZXZ[phibest]
\end{equation}
This result is consistent with $\phi=\shalf$ \cite{BY95,BWO99,BEG89}, 
and compares well with other estimates in the literature (for example, 
$\phi=0.501\pm 0.014$ in reference \cite{JvRR04}).

\begin{figure}[t]
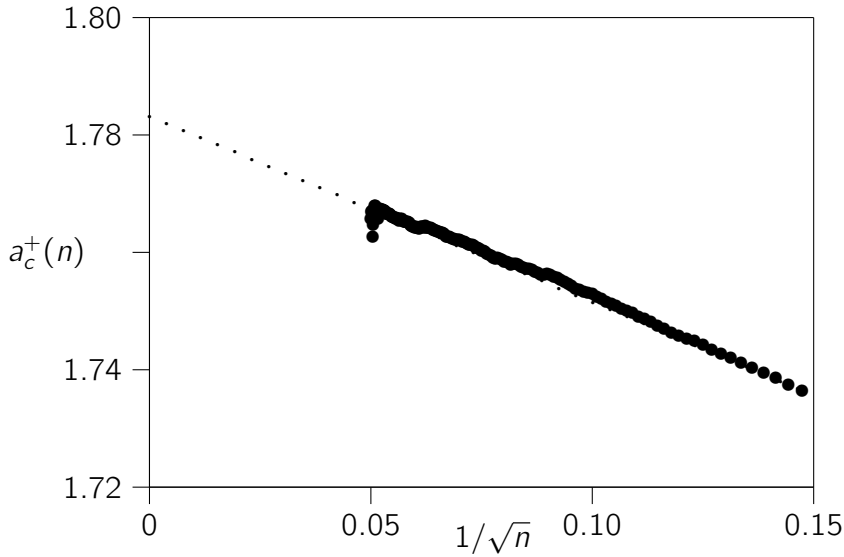

\input figure10-.tex
\caption{Extrapolating the intersections between specific heat curves
for  adsorbing walks in $\mathL_2^+$.
These data points are the intersections of the specific heat for
$n$ and $n+100$, for $n$ even and $n\in[44,400]$.  The data (except for
some data points at the largest values of $n$) line up
along a line if plotted against $1/\sqrt{n}$.  The location of the critical
point can be estimated by extrapolating the line to the left vertical axis;
this gives the estimate in equation \Ref{eqn29}.}
\label{figure5C}  %%ZXZ[figure5C]
\end{figure}

\subsubsection{The critical point $a_c^+$ and the specific heat $\C{C}_n(a)$:}

The best estimates above may be compard to estimates obtained from the
specific heat curves in figure \ref{figure5}.  These curves intersect each other
near $a_c^+$, and the region containing the intersections (in figure \ref{figure5B}) 
is magnified in figure \ref{figure5C}.

In general the location of the intersection between $\C{C}_n(a)$ and $\C{C}_m(a)$
is a function of $n$ and $m$.   The location of the critical point $a_c^+$
can  be estimated by extrapolating this dependence.  Consider
for example the intersections between the curves  
$\C{C}_n(a)$ and $\C{C}_{n+100}(a)$.  The location of these intersections
are plotted against $\sfrac{1}{\sqrt{n}}$ in figure \ref{figure5C},
where $n=2\, N$ and $N\in [23,200]$.  The data lie along a straight line, 
except for a few points at the largest values of $n$ (where the data is more
uncertain).  The best line through the data can be extrapolated to its intersection 
with the vertical axis (where $\sfrac{1}{\sqrt{n}} = 0$). This gives an estimate of  
$a_c^+$ as being located in the interval $[1.77,1.79]$.  Using a
linear least squares model for $n\geq 10$ gives the (extrapolated) estimate 
$a_c^+ \approx 1.7839$.  By examining the spread of the 
data in figure \ref{figure5C}, a confidence interval can be estimated. The result
is
\begin{equation}
a_c^+ = 1.784\pm 0.010 .
\label{eqn29}  %%ZXZ[eqn29]
\end{equation}
This estimate is consistent with the best estimate in equation \Ref{abest}.

An alternative approach to determining the critical point $a_c^+$, is
to consider the scaling of the specific heat in equation \Ref{eqn24EC}.
Taking ratios for $n$ and $m$, and then logarithms, give
\begin{equation}
\log \LB \Sfrac{\C{C}_n(a)}{\C{C}_m(a)} \RB
 = \alpha \phi \log \LB \Sfrac{n}{m} \RB
+ \log \LB \Sfrac{h_c(n^\phi(a\minus a_c^+))}{h_c(m^\phi(a\minus a_c^+))} \RB .
\label{eqn27}  %%ZXZ[eqn27]
\end{equation}
Observe that the last term is zero when $a=a_c^+$.   Since, in addition,
$\alpha = 2\minus \sfrac{1}{\phi}=0$ in this model,  this shows that an estimate
of the critical point is given by the solution of
\begin{equation}
\log \LB \Sfrac{\C{C}_n(a)}{\C{C}_m(a)} \RB = 0 .
\end{equation}
Solving this for $150\leq n \leq 500$ (and $n$ a multiple of $10$), 
and for $m=n\minus k$ where $k\in\{10,20,\ldots,100\}$ gives a large collection
of estimates of $a_c^+$, with mean
\begin{equation}
a_c^+ = 1.762 \pm 0.016 .
\label{eqn27AA}  %%ZXZ[eqn27AA]
\end{equation}
The confidence interval is one-half of difference between the maximum and
minimum estimates of $a_c^+$.  This estimate is slightly less than,but
still consistent with, the results in equation \Ref{abest} and 
equation \Ref{eqn29}.  Recall that it is also predicated on the assumption
that $\alpha=0$ (or $\phi = \sfrac{1}{2}$).

\begin{figure}[t]
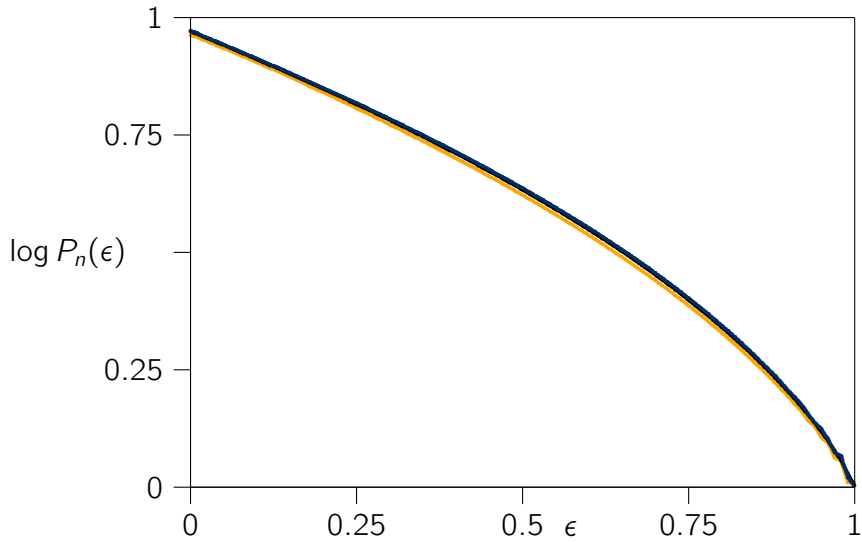

\input figure11-.tex
\caption{The microcanonical density function $P^+(\eps)$ of visits of adsorbing
square lattice walks.  These data the finite size approximations 
$P_n^+(\eps)$ to $P^+(\eps)$ for $n=100$ and $n=500$, as well as the
extrapolated estimate of $P^+(\eps)$.  The right derivative of $\log P^+(\eps)$
at $\eps=0$ is an estimate of the location of the critical adsorption point
$a_c^+$ in the model (see, for example, reference \cite{JvR15}).}
\label{figure6-B}  %%ZXZ[figure6-B]
\end{figure}

\subsubsection{The microcanonical density function:}
The microcanonical density function of visits in adsorbing positive walks is
determined from the microcanonical data in the model, and is given by
\begin{equation}
P^+(\eps) =\lim_{n\to\infty} (c_n^+(\lfl \eps n \rfl))^{1/n} = \lim_{n\to\infty} P_n^+(\eps),
\label{eqn59X}  %%ZXZ[eqn59X]
\end{equation}
where $P_n^+(\eps) = (c_n^+(\lfl \eps n\rfl))^{1/n}$ is a finite size approximation 
to $P^+(\eps)$. Existence of the limit is known (see for example reference \cite{JvR15}), 
and $\log P^+(\eps)$ is a concave function of $\eps$.  

$P^+(\eps)$ can be estimated by interpolating the finite size approximations
$P_n^+(\eps)$ and then extrapolating to $n=\infty$ by fitting a least squares
model to the data.  In figure \ref{figure6-B} the data for the extrapolated
function $P^+(\eps)$ is plotted together with $P_n^+(\eps)$ for $n=100$ and
$n=500$. 

A least squares fit of a quadratic to $\log P^+(\epsilon)$ for $\eps\in[0,0.1]$ 
gives the $\log P^+(\eps) \approx 0.97007\minus 0.58190\eps\minus 0.13030\eps^2$,
and by taking the right derivative and then taking $\eps\to 0^+$, an
estimate for the critical point is obtained:
\begin{equation}
a_c^+ \approx 1.789 .
\label{eqn27BB}  %%ZXZ[eqn27BB]
\end{equation}
This is close to the estimates obtained in equations \Ref{abest} and
\Ref{eqn29}, showing consistency in the data and the analysis above.

The free energy $\C{F}(a)$ is the Legendre transform of $\log P^+ (\epsilon)$.  
This may be estimated by fitting a polynomial to $\log P^+(\epsilon)$.  If a cubic
polynomial in $\eps$ is fitted to $\log P^+(\eps)$ for $0 \leq \eps \leq 0.5$,
then the estimated free energy for $a>a_c^+$ is approximately
\begin{equation}
\fl
F (a) \approx  1.0376 -0.1153\log a  -  (0.3306 \minus 0.5693\log a)
\sqrt{-1.3603 \plus 2.3422 \log a}.
\end{equation}
The critical point can be estimated as that location where the square root
in the above is zero.  This gives $a_c^+ \approx 1.787$.  Similarly, the
factor $(0.3306-0.5693\log a)$ vanishes when $a_c^+ \approx 1.787$.

\subsubsection{Metric data:}

\begin{figure}[t]
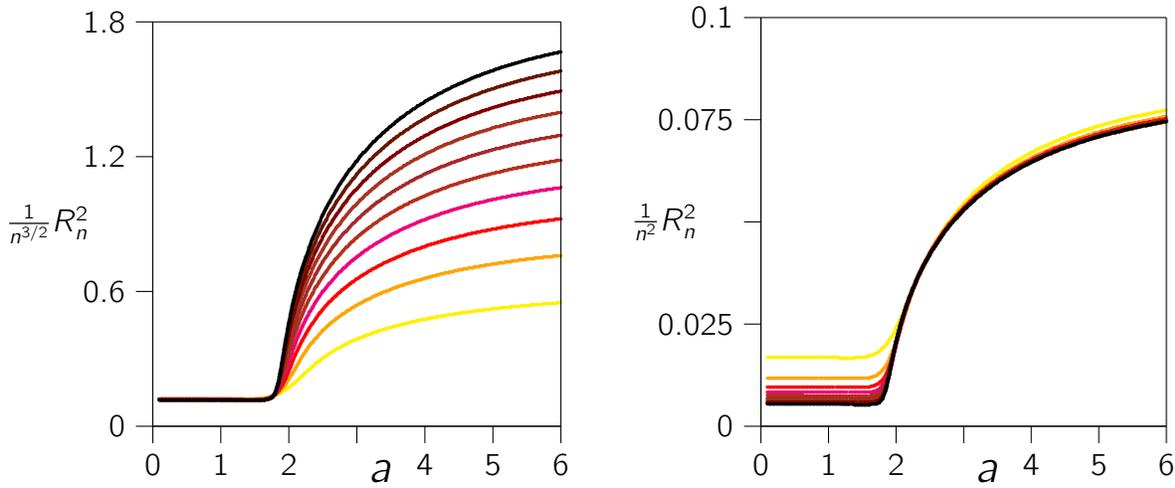

\input figure12-.tex
\caption{The mean square radius of gyration as a function of $a$,
for  adsorbing walks in $\mathL_2^+$.  In
the left panel $R^2_n$ is divided by $n^{2\nu}$.  For $a<a_c^+$ this
shows that $R^2_n \sim n^{2\nu}$, and for $a>a_c^+$ $R^2_n$ 
increases in size faster than $n^{2\nu}$ (since it is adsorbed, it
stays in the vicinity of the adsorbing boundary, and 
$R^2_n \sim n^2$ in this regime.  This is seen in
the right panel, where $R^2_n$ is divided by $n^2$.  For $a>a_c^+$
the curves collapse to a single underlying function, exposing the scaling
in the adsorbed phase.  The lengths of the walks varied
from $n=50$ (yellow) with colours increasing in hue to black when
$n=500$ in steps of $50$.}
\label{figure8}  %%ZXZ[figure8]
\end{figure}

The mean square radius of gyration $R_n^2$, and the mean height $H_n$
of the endpoint of the walk, are functions of the adsorption activity $a$.
In the desorbed phase (for $a<a_c^+$) it is expected that
$R_n^2 \sim n^{2\nu}$, and $H_n \sim n^\nu$, where $\nu = \sfrac{3}{4}$
is the metric exponent \cite{D89C}.  This scaling changes in the adsorbed phase
(when $a>a_c^+$) to $R_n^2 \sim n^2$ and $H_n \sim \hbox{constant}$.  
These expectations are confirmed by the data, as seen, for example, in figure 
\ref{figure8}, where data for the mean square radius of gyration are
normalised and then plotted as a function of $a$.  These graphs clearly 
show two scaling regimes, namely a high temperature phase 
(when $a< a_c^+$) where the walk has bulk critical exponents and is 
desorbed, and a low temperature phase where the walk stays near 
the adsorbing boundary and has critical exponents of a linear object.

\begin{figure}[t]
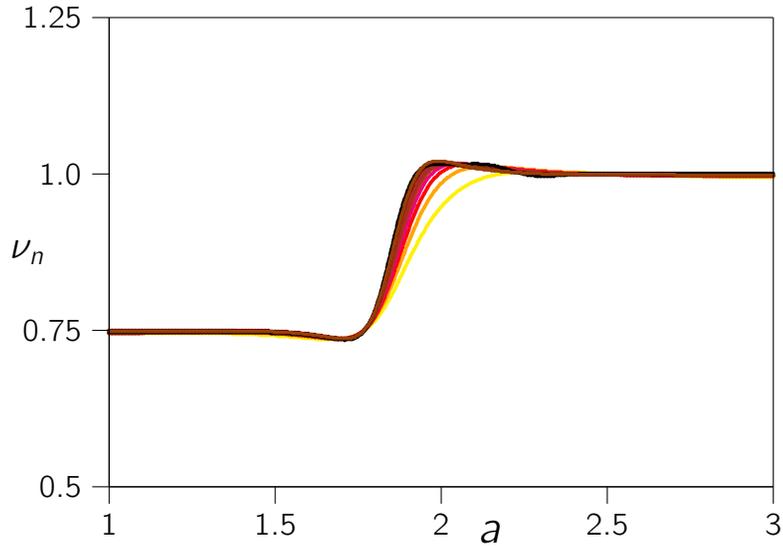

\input figure13-.tex
\caption{The metric exponent as a function of $a$, 
for  adsorbing walks in $\mathL_2^+$.  The exponent
was estimated from equation \Ref{eqn41} by fixing $n$ and then
averaging over $m$ for $100\leq m \leq 500$.  With increasing
$n$ the change from $\nu=\frac{3}{4}$ in the desorbed phase, to
$\nu=1$ in the adsorbed phase will become a step function at
$a=a_c^+$.  The lengths of the walks varied
from $n=50$ (yellow) with colours increasing in hue to black when
$n=500$ in steps of $50$.}
\label{figure8-nu}  %%ZXZ[figure8-nu]
\end{figure}

In general, the metric exponent associated with $R^2_n$ is a function
of $a$, and it will be denoted by $\nu_a$, where $\nu_a=\sfrac{3}{4}$
in the desorbed phase, and $\nu_a=1$ in the adsorbed phase.  This
exponent may be estimated from the mean square radius
of gyration $R_n^2$ data by examining the ratio
\begin{equation}
2\,\nu_{n,m}(a) = \frac{\log ( R_n^2/R_m^2 )}{\log (n/m)}.
\label{eqn41}   %%ZXZ[eqn41]
\end{equation}
Here, $\nu_{n,m}(a)$ is a function of $n$ and $m$.   By averaging over
$m$, the estimate $\nu_n(a) = \LA \nu_{n,m}(a)\RA_m$ may be determined.
In particular, fixing $n$ and taking the average over $m$ for $100 \leq m \leq 500$
in multiples of $5$ (and for $m\not=n$) gives estimates of $\nu_n(a)$.
These results are plotted in figure \ref{figure8-nu} for $n\in\{50,100,150,\ldots,500\}$.
The data for $a\leq 1.5$ gives $\nu \approx 0.747$, and for $a \geq 1.95$,
$\nu \approx 1.01$.  These results are evidence for the exact value 
$\nu_a = \sfrac{3}{4}$ in the desorbed phase, and $\nu_a=1$ in the adsorbed
phase.

\begin{figure}[b]
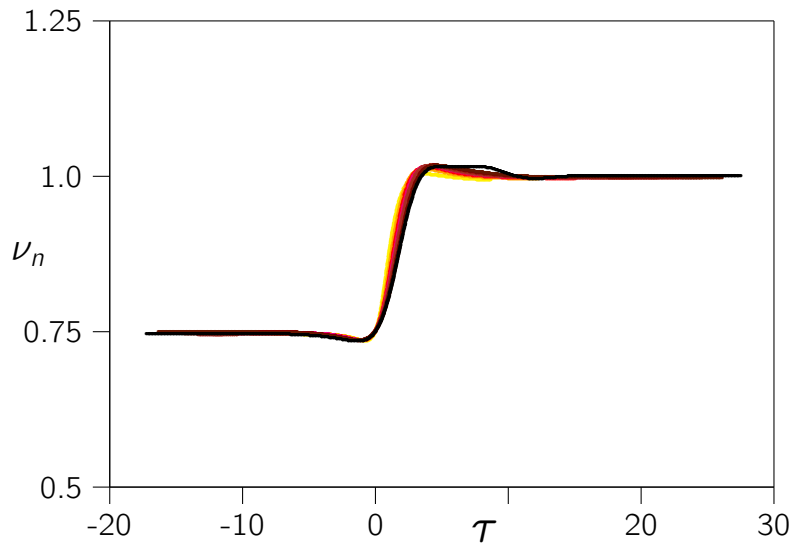

\input figure14-.tex
\caption{The same data as in figure \ref{figure8-nu}, but now plotted
against the rescaled variable $\tau = n^\phi(a-a_c^+)$,
for  adsorbing walks in $\mathL_2^+$.  This rescaling
of the activity $a$ collapses the curves to a single underlying curve
characterising the metric scaling of the walk in the desorbed and
adsorbed phases.  The lengths of the walks varied
from $n=50$ (yellow) with colours increasing in hue to black when
$n=500$ in steps of $50$. }
\label{figure6-D}  %%ZXZ[figure6-D]
\end{figure}

The function $\nu_n(a)$ should scale with the combined variable
$\tau = n^\phi(a\minus a_c^+)$.  That is, one may expect that
$\nu_n (a) = \hbox{\Large$\nu$}(\tau)$, where \hbox{\Large$\nu$}
is a scaling function.  In figure \ref{figure6-D} the data in figure
\ref{figure8-nu} are rescaled by plotting against $\tau$ to
uncover the scaling function \hbox{\Large$\nu$}.

\begin{figure}[t]
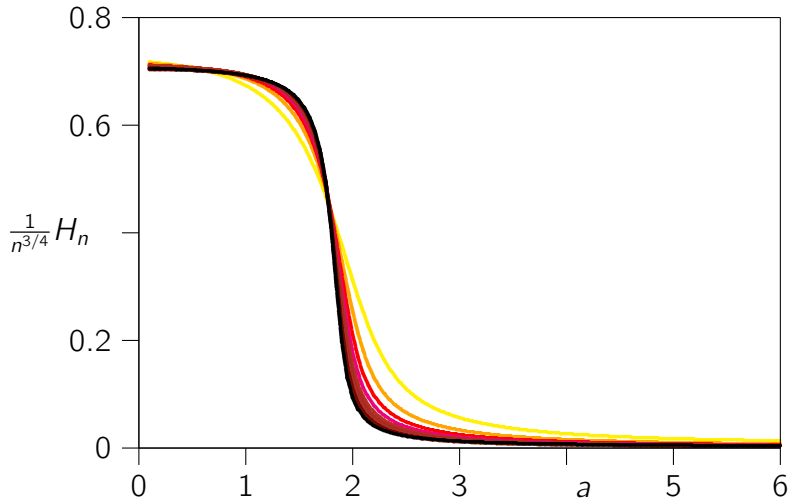

\input figure15-.tex
\caption{The mean height $H_n$ of the endpoint 
of adsorbing walks in $\mathL_2^+$
of length $n$ as a function of $a$.  On the left $n^{-\nu}H_n$ is plotted
as function of $a$.  This quantity decreases sharply as $a$ passes through
the adsorption critical point $a_n^+$.   The lengths of the 
walks varied from $n=50$ (yellow) with colours increasing in hue to black when
$n=500$ in steps of $50$. }
\label{figure8HH}  %%ZXZ[figure8HH]
\end{figure}

The (normalised) average height of the endpoint of the walk is plotted as a function
of $a$ in figure \ref{figure8HH}.   These data show a clear transition
where the scaling of $H_n$ changes.  That is, the metric exponent associated 
with $H_n$ is a function of $a$, and is denoted by $\nu_a^\perp$, 
where $\nu_a^\perp = \nu= \sfrac{3}{4}$ if $a<a_c^+$, and 
$\nu_a^\perp = 0$ if $a>a_c^+$, so that $H_n \sim n^{\nu_a^\perp}$.
The graph of $n^{-3/4}H_n$ contains a set of curves which decreases with
increasing $a$.  These curves intersect each other close to $a_c^+$.

\begin{figure}[b]
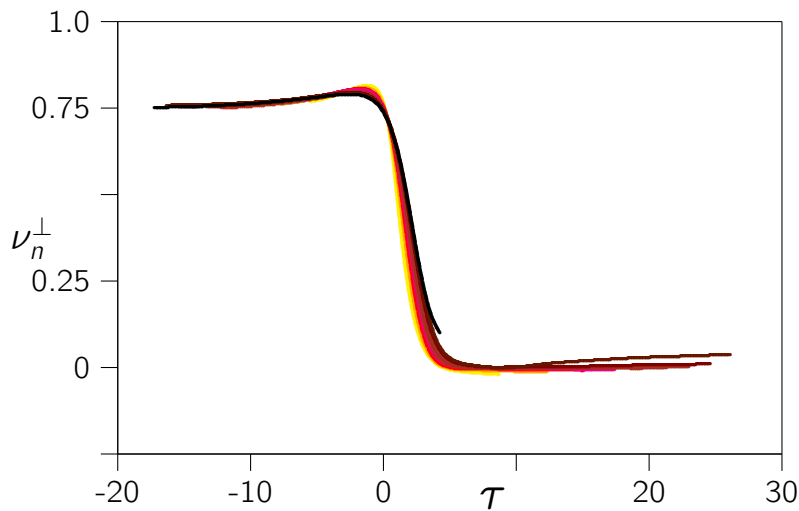

\input figure16-.tex
\caption{The vertical metric exponent $\nu^\perp$ as a function of $a$
for  adsorbing walks in $\mathL_2^+$.  
The exponent was estimated from equation \Ref{eqn42} by fixing $n$ and 
then averaging over $m$ for $100\leq m \leq 500$ .  With increasing
$n$ the exponent changes from $\nu^\perp=\nu=\frac{3}{4}$ in the desorbed 
phase, to $\nu^\perp=0$ in the adsorbed phase.  Data for $n=500$ and 
$\tau>5$ did not converge, and is left out of this figure.  The lengths of the 
walks varied from $n=50$ (yellow) with colours increasing in hue to black when
$n=500$ in steps of $50$. }
\label{figure6-nuH}  %%ZXZ[figure6-nuH]
\end{figure}

An approach similar to the ratio method
in equation \Ref{eqn41} may be used to estimate $\nu_a^\perp$ (which will be
referred to as the \textit{vertical metric exponent}):
\begin{equation}
\nu^{\perp}_{n,m} (a) = \frac{\log (H_n/H_m)}{\log (n/m)}.
\label{eqn42}   %%ZXZ[eqn42]
\end{equation}
This estimate of $\nu_a^\perp$ is a function of $n$ and $m$, and may be
averaged over $m$ to obtain $\nu^{\perp}_n(a) = \LA \nu_{n,m}(a)\RA_n$.
Taking the average for $100\leq m \leq 500$ in multiples of $5$ to estimate
$\nu^\perp_n(a)$ gives the curves in figure \ref{figure6-nuH} when plotted
against the combined variable $\tau=n^\phi(a\minus a_c^+)$. 
It is seen in the graph that if $a<a_c^+$, then $\nu_n (a) \approx \sfrac{3}{4}$,
but for $a>a_c^+$, $H_n \simeq \hbox{const}$ so that 
$\nu_n^{\perp}(a) \approx 0$. 

\begin{figure}[t]
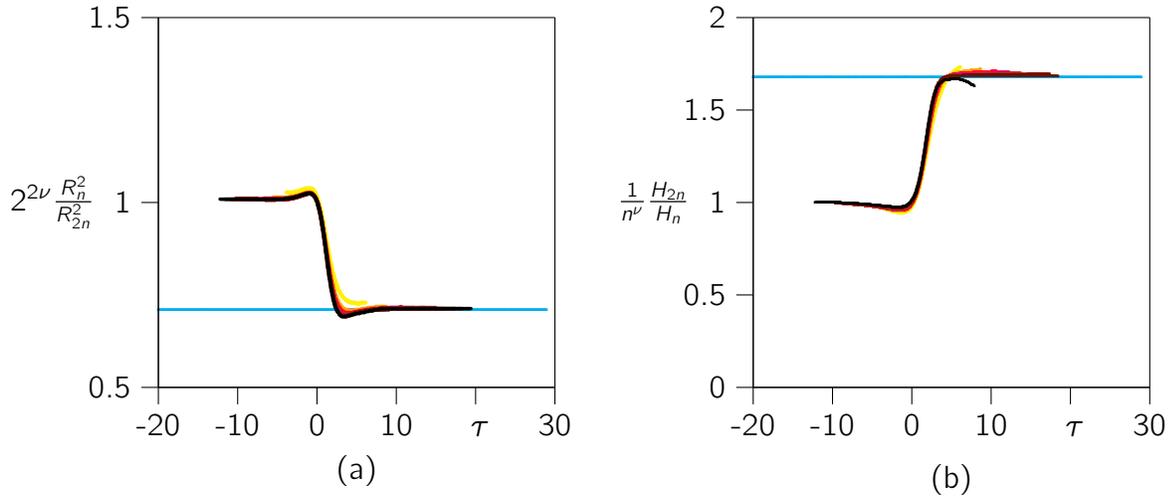

\input figure17-.tex
\caption{Metric scaling for adsorbing walks in $\mathL_2^+$.
(a) The ratio $R^2_{2n}/(n^{2\nu}$of the mean square radius of 
gyration plotted as a function of $\tau=n^\phi(a-a_c^+)$
is displayed on the left.  The ratio is given in equation \Ref{eqn58}
and the data shows that it is approximately equal to $1$ in the desorbed phase
and approximately equal to $2^{3/2 - 2} \approx 0.71$ in the adsorbed phase.
(b) The ratio $H_{2n}/(n^\nu H_n)$ plotted as a function of $\tau$.
This ratio is defined in equation \Ref{eqn59} and is approximately equal
to $1$ in the desorbed phase, and equal to $2^{3/4}$ in the adsorbed
phase.  The data in both panels are for $n\in\{25,50,75,100,\ldots,250\}$
and increase in hue from yellow ($n=25$) to black ($n=250$).
}
\label{figure8RR}  %%ZXZ[figure8RR]
\end{figure}

Since $R^2_n \sim n^{2\nu_a}$ where $\nu_a = \nu = \sfrac{3}{4}$
if $a<a_c^+$, and $\nu_a = 1$ if $a>a_c^+$, ratios of $R^2_n$ may be defined by
\begin{equation}
\frac{R^2_{n}}{R^2_{2n}} \approx 2^{-2\nu_a} ,\q\hbox{or}\q
\frac{2^{2\nu}R^2_{n}}{R^2_{2n}} \approx 
\cases{
1, & if $a< a_c^+$; \\
2^{2\nu - 2\nu_a} = 2^{-1/2} , & if $a>a_c^+$.
}
\label{eqn58} %%ZXZ[eqn58]
\end{equation}
In figure \ref{figure8RR} the quantity $\sfrac{2^{2\nu}R_{n}^2}{R_{2n}^2}$
is plotted as a function of the rescaled variable $\tau = n^{1/2}(a\minus a_c^+)$ 
for $n$ from $25$ to $250$ in steps of $25$.  In the desorbed phase this ratio
should be equal to $1$, but, in the adsorbed phase, it should be equal to
$2^{1.5-2.0} \approx 0.71$.  This is clearly seen in the graph.  The curves coincide
well over the entire range of $n$ and $\tau$, and decreases sharply close 
to the critical adsorption point at $\tau=0$.  A similar approach using the 
heights of the endpoint involves the ratios
\begin{equation}
\frac{H_{n}}{H_{2n}} \approx 2^{-\nu_a^\perp} ,\q\hbox{or}\q
\frac{2^\nu H_{n}}{H_{2n}} \approx 
\cases{
1, & if $a< a_c^+$; \\
2^{\nu-\nu^\perp}  = 2^{3/4}, & if $a>a_c^+$.
}
\label{eqn59} %%ZXZ[eqn59]
\end{equation}
The quantity $\sfrac{2^\nu H_{n}}{H_{2n}}$  is plotted against $\tau$
on the right panel in figure \ref{figure8RR}.  In the desorbed phase
this ratio should be equal to $1$, as seen in the graph.  In adsorbed phase
the scaling of $H_n$ changes, and the ratio should be equal to 
$2^{3/4} \approx 1.68$, as seen in the graph.

\begin{figure}[t]
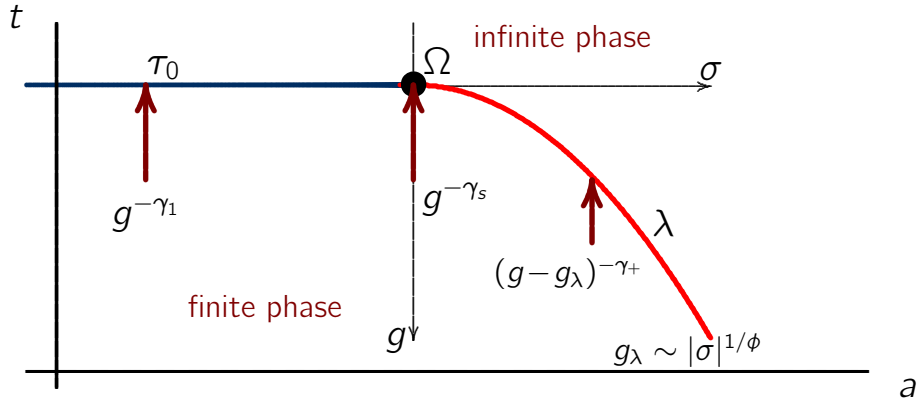

\input figure18-.tex
\caption{The phase diagram of the generating function $G(a,t)$.  
The radius of converge of $G(a,t)$ is denoted by $t_c^+(a)$, and it is
the \textit{critical curve}.  This curve is divided into two parts
(a transition to a low temperature phase marked by $\tau_0$ and 
a transition to a high temperature phase marked by $\lambda$).
The $\taus0$ and $\lambda$ curves meet  
in the \textit{critical point} $\Omega$
(located at the critical adsorption transition $a_c^+$).   The critical
curve separates a ``finite phase" (where $G(a,t)$ is convergent and
dominated by walks of finite length) from an ``infinite phase" (where
$G(a,t)$ is divergent).  The scaling of $G(a,t)$ is described by setting
up a coordinate system $(g,\sigma)$ with origin in $\Omega$, where
$\sigma = (a{-}a_c^+)$ and $g=\frac{1}{\mu_2} {-} t$.  The scaling of $G(a,t)$ 
on approach to the critical curve is given by equation \Ref{eqn46},
depending on whether $\sigma<0$, $\sigma=0$, or $\sigma>0$.}
\label{figure4-6}  %%ZXZ[figure4-6]
\end{figure}

\subsection{The generating function}

The generating function is given by the series
\begin{equation}
G (a,t) = \sum_{n=0}^\infty \sum_{v=0}^n c_n^+(v)\,a^vt^n .
\label{eqn45}  %%ZXZ[eqn45]
\end{equation}
Approximations of $G(a,t)$ is given by the truncated sum
\begin{equation}
G_N(a,t) = \sum_{n=0}^N \sum_{v=0}^n c_n^+(v)\,a^vt^n ,
\label{eqn46a}  %%ZXZ[eqn46a]
\end{equation}
where $c_n^+(v)$ is approximately enumerated by the GAS algorithm.
In this study, $G_{500}(a,t)$ was estimated using the approximate values of
$c_n^+(v)$ obtained in the simulations.

\begin{figure}[t]
\begin{center}
\includegraphics[scale=0.6]{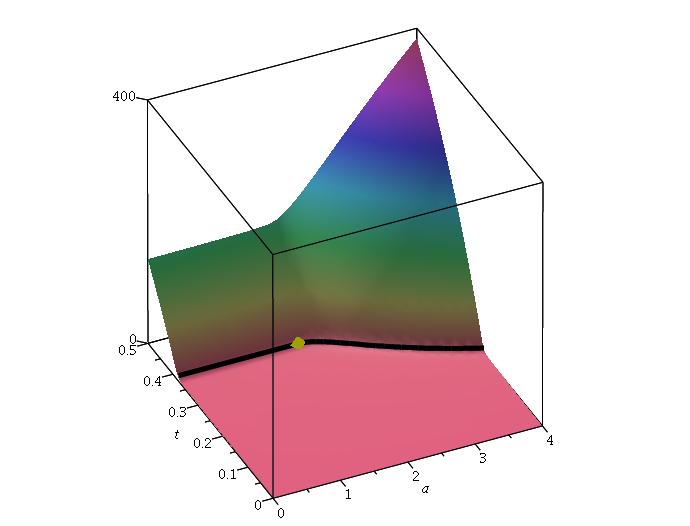}
\end{center}
\caption{A plot of $\log G_{500}(a,t)$ as a function of $(a,t)$ 
for adsorbing walks in $\mathL_2^+$.  The
critical curve is the black curve along the surface, and the critical 
point separating the $\tau_0$ and $\lambda$ curves (see figure
\ref{figure4-6}) is denoted.  For $t<t_c^+(a)$ the generating
$G(a,t)$ is finite, and for $t>t_c^+(a)$ it is divergent.  These two regimes
are clearly visible.}
\label{figureAdsrb-2}   %%ZXZ[figureAdsrb-2]
\end{figure}

The \textit{critical curve} $t_c^+(a)$ of $G(a,t)$ is its radius of convergence as a 
function of $a$.  By equation \Ref{eqn25F} and by equation \Ref{eqnZ},
\begin{equation}
\log t_c^+(a) = - \C{F}(a) = - \lim_{n\to\infty} \sfrac{1}{n} \log  Z_n(a) .
\label{eqn46A}  %%ZXZ[eqn46A]
\end{equation}
The critical curve is shown schematically in figure \ref{figure4-6}.
$G(a,t)$ is singular when $t=t_c^+(a)$, and if $t>t_c^+(a)$, then $G(a,t)$ is divergent.

The critical curve is parametrized by the scaling fields $(\sigma,g)$ as shown
in figure \ref{figure4-6}.  The critical point when $a=a_c^+$  is located
at $(a_c^+,t_c^+)$, where $t_c^+\equiv t_c^+(a_c^+)$, and it divides the critical curve
$t_c^+(a)$ into two parts.  The part marked by $\tau_0$ corresponds to
a transition to desorbed walks, so that $t_c^+(a)=1/\mu_2$ in this regime
(which is a transition to a high temperature phase).
For $a>a_c^+$ the approach to $t_c^+(a)$ is to adsorbed walks, along the
critical curve marked by $\lambda$ (which is a transition to a low temperature phase).  

The phase diagram in figure \ref{figure4-6} may be described in terms of
the coordinates $g=(\sfrac{1}{\mu_2} \minus t)$ and $\sigma = (a\minus a_c^+)$.  
The behaviour of $G(a,t)$ along its singular points along the critical curve
is  described by 
\begin{equation}
G(a,t) \sim \cases{
g^{-\gamma_1}, & \hbox{along $\tau_0$;} \\
g^{-\gamma_s}, & \hbox{at the critical point $a=a_c^+$;}\\
g^{-\gamma_+}, & \hbox{along $\lambda$.}
}
\label{eqn46}  %%ZXZ[eqn46]
\end{equation}
In two dimensions exact values are known for the exponents:
The exponent $\gamma_1 = \sfrac{61}{64}$ \cite{C87} 
is the entropic exponent of half-space walks,
and $\gamma_s = \sfrac{93}{64}$ \cite{BEG89}
is the surface exponent of adsorbing half-space walks at the 
critical point $a_c^+$.  The exponent $\gamma_+$
is the entropic exponent of adsorbed walks, and is given by the
entropic exponent of self-avoiding walks in one dimension lower.
In one dimension, this is $\gamma_+=1$.

A plot of $G_N(a,t)$ is shown in figure \ref{figureAdsrb-2}.  The horizontal
plane is the $(a,t)$-plane, and the critical curve in figure \ref{figure4-6}
is shown as a black curve with the critical point shown.  Below the critical
curve $G(a,t)$ is finite, and approximated well by $G_N(a,t)$ (for large
values of $N$ not too close to the critical curve).  Above the critical curve
$G(a,t)$ is divergent, while $G_N(a,t)$, which is a polynomial, is finite.

\begin{figure}[t]
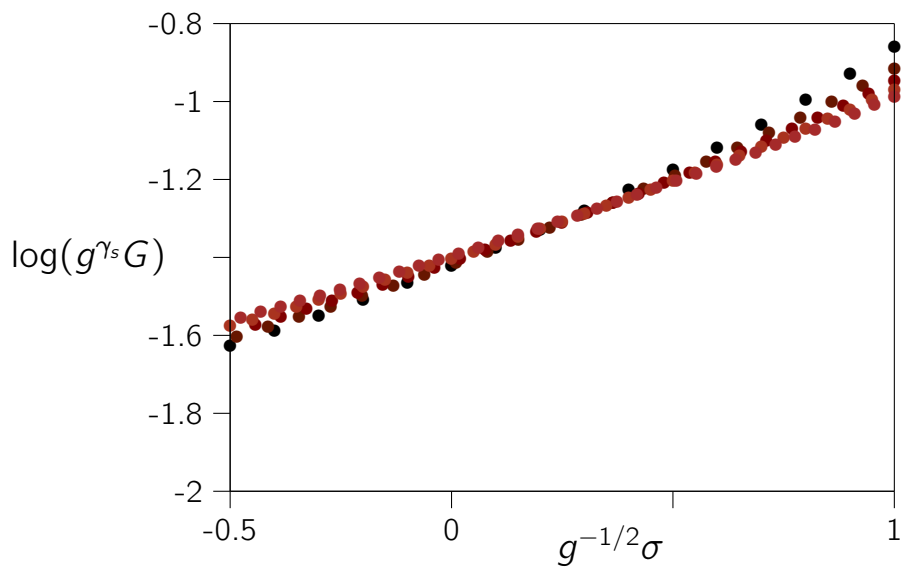

\input figure20-.tex
\caption{Scaling of the generating function near the critical adsorption
point for adsorbing walks in $\mathL_2^+$.  
The data points above are collected for $g\in [0.01,0.05]$ and
$-\frac{1}{2}\sqrt{g}\leq \sigma \leq \sqrt{g}$. The surface exponent $\gamma_s
=\frac{93}{64}$, its exact value.}
\label{figureGscale}  %%ZXZ[figureGscale]
\end{figure}

The exponent $\gamma_1$ can be estimated by putting $a=1$ so that
$G(1,t) \sim g^{-\gamma_1}$ where $g=(t_c^+(1)\minus t)$ and 
$t_c^+(1) = \sfrac{1}{\mu_2}$ (and where $\mu_2$ is the growth constant of the
walks in two dimensions).  Thus, estimate $\gamma_1$ by noting that
\begin{equation}
\frac{\log G(1,t)}{\log g} = - \gamma_1 + \Sfrac{C_1}{\log g} + \Sfrac{C_2}{\log^2 g}
+ \ldots .
\label{eqn47}  %%ZXZ[eqn47]
\end{equation}
Proceed by approximating $G(1,t)$ by $G_N(1,t)$ (with $t < t_c^+(1)$ so that
$G_{500}(1,t)$ is a good approximation of $G(1,t)$).  A least squares fit of the
ratio on the left to a quadratic in $\sfrac{1}{\log g}$ gives the estimate
\begin{equation}
\gamma_1=0.952\ldots .
\end{equation}
This result is very close to the exact value $\sfrac{61}{64} = 0.953\ldots$.
Similar analysis for $a>1$ and $a<a_c^+$ gives results slightly larger, since
the critical point at $a_c^+$ influences the data in its vicinity for finite
values of $N$.

\begin{figure}[b]
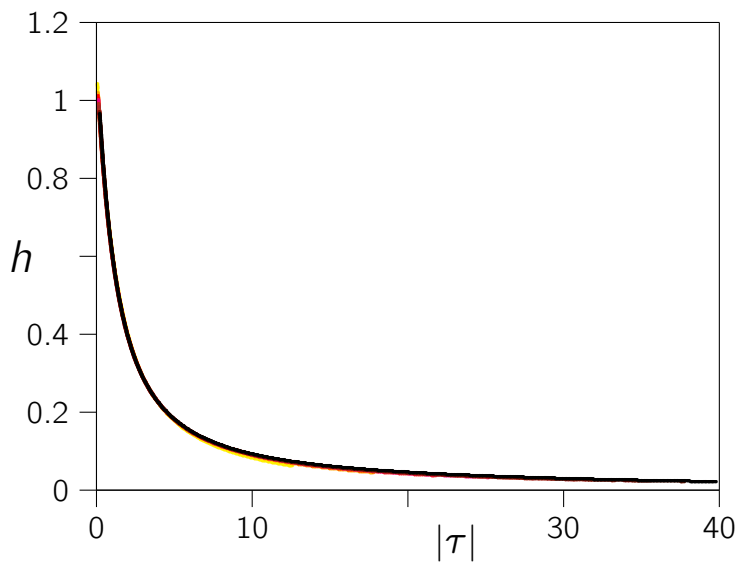

\input figure21-.tex
\caption{This figure verifies the scaling of the partition function
for adsorbing walks in $\mathL_2^+$ as proposed
in equation \Ref{eqnZscaling}.  The curves are plots of 
$n^{-29/64}Z_n(a)\,\mu_2^{-n}$ against the combined variable
$|\tau | = n^{1/2}|a\minus a_c^+|$ (with the critical point approximated 
by $a_c^+ =1.78$ and $a<a_c^+$).  The curves are walks of
lenghts $n\in\{50,100,150,\ldots,500\}$.  All the data collapse to
expose the scaling function $h$ in equation \Ref{eqnZscaling}.  
The lengths of the walks varied
from $n=50$ (yellow) with colours increasing in hue to black when
$n=500$ in steps of $50$. }
\label{figureZscale}  %%ZXZ[figureZscale]
\end{figure}

A similar analysis with $a=a_c^+$ gives the estimate
\begin{equation}
\gamma_s=1.429\ldots,
\end{equation}
for the surface exponent at the critical adsorption point.  This is close in
value to the exact result $\sfrac{93}{64} = 1.453\ldots$.

The case that $a>a_c^+$ may also be analysed.  The adsorbed walk 
should have the statistics of the self-avoiding walk in one dimension, so that
the entropic exponent is $\gamma_+ = 1$ (this is the value of the
exponent $\gamma$ in one dimension).  Putting $a=3.5$ and plotting 
$Z_n^{1/n}(3.5)$ against $n$ gives an estimate for the critical value 
$t_c^+(3.5)$.  In this case the data quickly converges to $t_c^+(3.5)=0.260\ldots$ 
(to three decimal places).  Assuming that $t_c^+(3.5)=0.260$ and choosing 
the scaling field $g=(0.260 \minus t)$ gives the model
\begin{equation}
\Sfrac{\log G(3.5,t)}{\log g} = - \gamma_+ + \Sfrac{C_1}{\log g} + \Sfrac{C_2}{\log^2 g}
+ \ldots ,
\label{eqn48}  %%ZXZ[eqn48]
\end{equation}
similar to the above, but now with the exponent $\gamma_+$.   Plotting
the left hand side as a function of $\kappa=\sfrac{1}{\log g}$, and fitting the data to
a quadratic in $\kappa$, give the estimate $\gamma_+ = 1.00\ldots$.  There remains, 
however, some curvature in the model for small values of $g$ (and of $\sigma$),
so that there may remain strong systematic corrections to this result (however,
$\gamma_+ = 1$ is consistent with this result).

In the vicinity of the critical point $(a_c^+,t_c^+)$ the generating function
should exhibit scaling given by
\begin{equation}
G(a,t) \sim g^{-\gamma_s} \,f(g^{-\phi} \sigma),
\label{eqnGscale}   %%ZXZ[eqnGscale]
\end{equation}
where $\phi=\sfrac{1}{2}$ is the crossover exponent and $f$ is a scaling
function.  That is, plotting $g^{\gamma_s}\,G(a,t)$ against the combined 
variable $g^{-\phi} \sigma$ should expose the scaling function $f$. 
In figure \ref{figureGscale} this is done by plotting
$\log (g^{\gamma_s}\,G(a,t) )$ against $g^{-1/2}\sigma$ for 
$g=\sfrac{1}{\mu_2}\minus t \in [0.01,0.05]$ and $\sigma=a\minus a_c^+ 
\in [-\sfrac{1}{2}g^{1/2},g^{1/2}]$.

The partition function (see equation \Ref{eqnZ}) should also exhibit
scaling for large $n$, given by
\begin{equation}
Z_n(a) \sim n^{\gamma_t-1}\,h(n^\phi (a_c^+ \minus a))\,\mu_a^n,
\label{eqn53}   %%ZXZ[eqn53]
\end{equation}
where $\log \mu_a = \C{F}(a)$ and $\log \mu_a = - \log t_c^+(a)$.
The exponent $\gamma_t$ can be related to the $\gamma$-exponents
as $a\to a_c^+$ (that is, as $\sigma\to 0$), and namely to the surface exponent $\gamma_s$,
where the walk is critical with respect to the adsorption transition.
By noting that $G(a,t) = \sum_n Z_n(a)\,t^n$, and approximating
the summation by an integral, it follows that $G(a,t) \sim
g^{-\gamma_t} h(0)$ if $a=a_c^+$ and $g=t_c^+\minus t$.
This shows that
\begin{equation}
\gamma_t-1 = \gamma_s - 1 = \Sfrac{93}{64} - 1 = \Sfrac{29}{64}.
\label{eqn71g}   %%ZXZ[eqn71g]
\end{equation}
That is, when $a$ is close to $a_c^+$ (and $a<a_c^+$), 
the partition function has asymptotic behaviour
\begin{equation}
Z_n(a) \sim n^{29/64} h(n^\phi(a_c^+\minus a))\, \mu_a^n .
\label{eqnZscaling}   %%ZXZ[eqnZscaling]
\end{equation}
This result may be tested by plotting
$n^{-29/64} Z_n(a) \, (t_c^+(a))^n$ against $|\tau | = n^{1/2} |a\minus a_c^+|$.
All the data should collapse to the same universal curve exposing the
scaling function $h$.  This is done with $a<a_c^+$ (and $t_c^+(a) = \sfrac{1}{\mu_2}$)
in figure \ref{figureZscale} for $n\in\{50,100,150,\ldots,500\}$ and 
$0 \leq a < a_c^+$.

\begin{figure}[t]
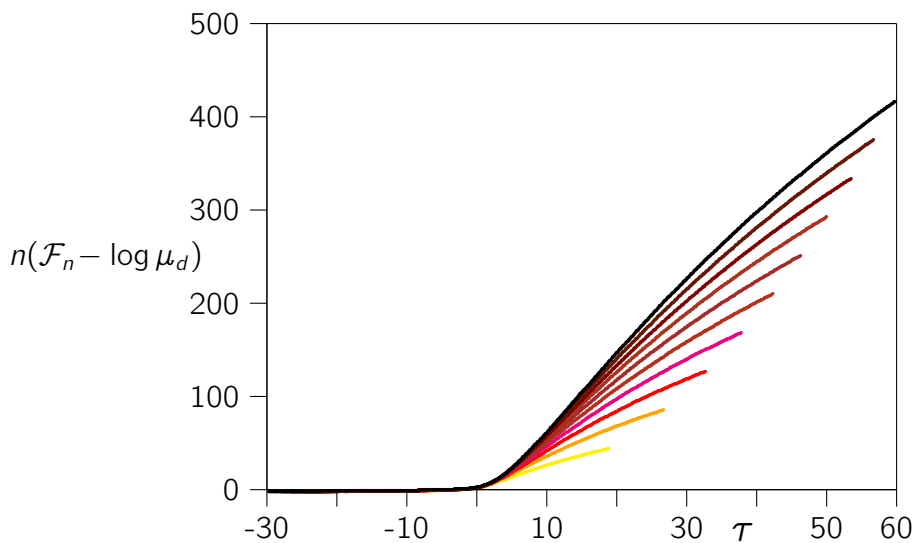

\input figure22-.tex
\caption{The scaled free energy $n(\C{F}_n(a) - \log \mu_d)$ as a function
of $\tau=n^{1/2}(a-a_c^+)$ for  adsorbing walks in $\mathL_3^+$.  
The lengths of the walks varied
from $n=50$ (yellow) with colours increasing in hue to black when
$n=500$ in steps of $50$.   The data
collapse for small $|\tau|$ (this is the critical scaling regime), 
and will approach a limiting curve as $n\to\infty$, also for large $\tau$,
given by the limiting free energy.}
\label{figure4B-3}  %%ZXZ[figure4B-3]
\end{figure}

\subsection{Adsorbing walks in the cubic lattice}

The (finite size) free energy $\C{F}_n(a)$ is a function of the 
combined variable $\tau=n^{1/2}(a\minus a_c^+)$ (see equations 
\Ref{eqn24} and \Ref{eqn26}; note that $\phi=\sfrac{1}{2}$ for
adsorbing walks in three dimensions \cite{HG94}).  Plotting 
$\C{F}_n(a)$ against $\tau$ for data in the cubic lattice gives a
graph similar to figure \ref{figure4B}.  In figure \ref{figure4B-3}
the scaled free energy $n(\C{F}_n(a)\minus\log\mu_3)$ 
is plotted against $\tau$.  The data shows a clear transition in the
model from a desorbed to an adsorbed phase.

\begin{figure}[b]
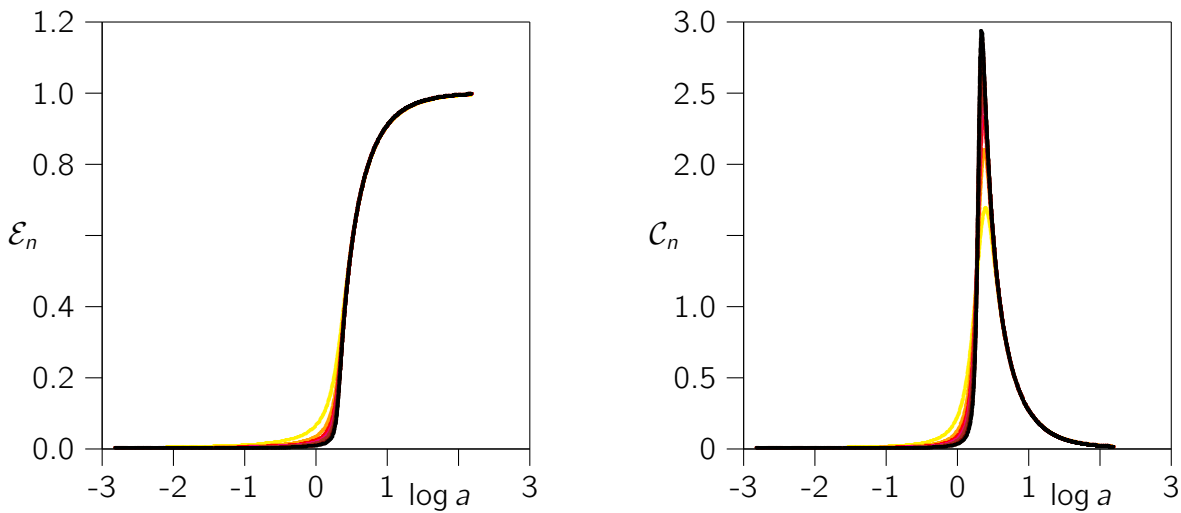

\input figure23-.tex
\caption{The energy $\C{E}_n(a)$ and specific heat $\C{C}_n(a)$
(see equation \Ref{eqn24EC}) as a function of $\log a$ for
adsorbing walks in the cubic lattice.  See figure \ref{figure5} for
the similar results in the square lattice.  The lengths of the walks varied
from $n=50$ (yellow) with colours increasing in hue to black when
$n=500$ in steps of $50$.}
\label{figure1-3a}  %%ZXZ[figure1-3a]
\end{figure}

\begin{figure}[t]
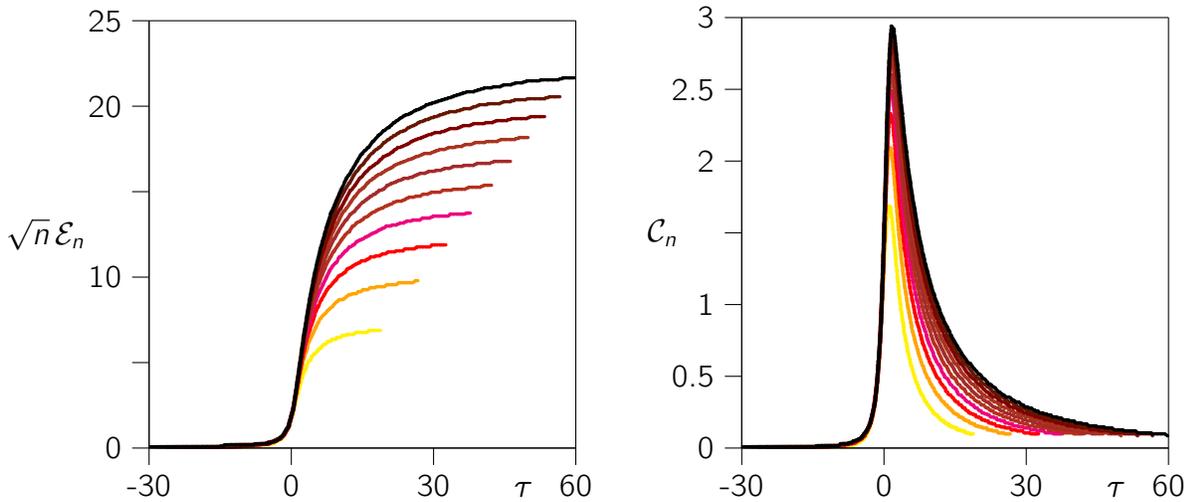

\input figure24-.tex
\caption{The scaled energy density $\sqrt{n}\C{E}_n(a)$ (left) and the
specific heat $\C{C}_n(a)$ (right) for  adsorbing walks in $\mathL_3^+$
as a function of the $\tau = 
n^{1/2}(a-a_c^+)$ for $n$ in steps of $50$ to $500$.  The lengths of the 
walks varied from $n=50$ (yellow) with colours increasing in hue to black when
$n=500$ in steps of $50$.  The data
collapse for small values of $|\tau|$ (this is the critical scaling regime
which contains the critical point).  For larger values of $|\tau|$ the
curves will approach a limiting curve as $n\to\infty$
given by the limiting free energy.}
\label{figure4-3a}  %%ZXZ[figure4-3a]
\end{figure}

Derivatives of the free energy to $\log a$ gives the (finite size) energy 
density $\C{E}_n(a)$ and (finite size) specific heat $\C{C}_n(a)$. These are plotted
in figure \ref{figure1-3a} against $\log a$ and 
in figure \ref{figure4-3a} against $\tau$.  In these plots, as in 
figure \ref{figure4B-3}, the critical point was approximated by
$a_c^+ = 1.31$.   This is a close approximation of the best estimate for 
the critical point from our data (see equation \Ref{abest3}).

\subsubsection{Location of the critical point $a_c^+$:}
The location of the critical adsorption point can be determined using
the same analysis as in section \ref{section311}, and in particular
using equation \Ref{eqn44C} as a starting point.  That, is for 
given values of $n$ and $m$, an estimate $a_{n,m}^+$
of $a_c^+$ can be obtained by solving for $a$ in 
\begin{equation}
\hspace{-1cm}
\frac{\log (n\,\C{E}_n(a))}{\log (m\, \C{E}_m(a))}\frac{\log m}{\log n}
= 1.
\end{equation}
Here, the choice $m=n\minus 100$ worked well, and $n$ was assigned 
values starting at $n=200$ to $n=500$ in steps of $1$.  The estimates 
$a_n^+ \equiv a_{n,n-100}^+$ showed a dependence on $n$, systematically
decreasing with increasing $n$.  The best estimate is obtained by using the
model  $a_n^+ = a_c^+ - \sfrac{c_1}{\log^2 n}$ suggested by equation
\Ref{eqn44D}.   A least squares fit for all $n\in [200,500]$ gives the
best estimate
\begin{equation}
a_c^+ = 1.3055 \pm 0.0061 .
\label{abest3}   %%ZXZ[abest3]
\end{equation}
The confidence interval is obtained by doubling the square root of
the variance of the estimates $a_n^+$.  This result is slightly smaller than
the result in reference \cite{JvRR04}, namely $a_c^+ = 1.334\pm 0.027$ 
(obtained by using a Multiple Markov Chain implementation of the 
Berretti-Sokal algorithm \cite{BS85}).

The estimate \Ref{abest3} can be used to determine the crossover exponent
$\phi$.  This is again done by considering the scaling of the specific heat 
(equation \Ref{eqn24EC}).  It is expected that $\C{C}_n(a_c^+) 
\sim n^{\alpha\phi} h_c(0)$ when $a=a_c^+$.  An estimate of 
$\alpha\phi$ is obtained by computing $\sfrac{\log \C{C}_n(a_c^+)}{\log n}$
for a range of values of $n$ (in this case $100 \leq n \leq 500$).  The average
is taken as the best estimate and a confidence interval is estimated by
doubling the square root of the variance of the estimates.  This
gives $\alpha\phi = 0.0106 \pm 0.0116$.  Determine the best estimate for
$\phi$ by using equation \Ref{eqn25}:
\begin{equation}
\phi = 0.5053 \pm 0.0053 .
\label{phibest3}  %%ZXZ[phibest3]
\end{equation}
This result compares well with the estimate $\phi=0.5005\pm 0.0036$ 
for adsorbing walks in reference \cite{JvRR04}.

\subsubsection{The critical point $a_c^+$ and the specific heat $\C{C}_n(a)$:}
\begin{figure}[t]
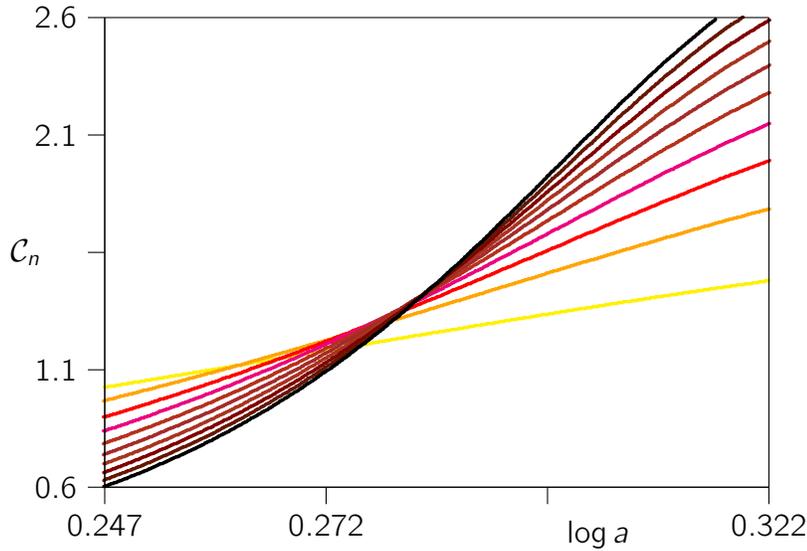

\input figure25-.tex
\caption{The specific heat curves in figure \ref{figure1-3a} 
for  adsorbing walks in $\mathL_3^+$ intersect
each other close to the critical adsorption point.  This figure is a magnification of
the curves close to where they intersect.  See figure \ref{figure5B} for the
similar results in the square lattice.  The lengths of the walks varied
from $n=50$ (yellow) with colours increasing in hue to black when
$n=500$ in steps of $50$.}
\label{figure2-3a}  %%ZXZ[figure2-3a]
\end{figure}

The best estimate for $a_c^+$ above (see equation \Ref{abest3}) should be
examined by comparing it to estimates obtained from the specific heat
curves in figure \ref{figure1-3a}.  These curves intersect each other
near $a_c^+$, and the region containing the intersections
is magnified in figure \ref{figure2-3a}.

\begin{figure}[t]
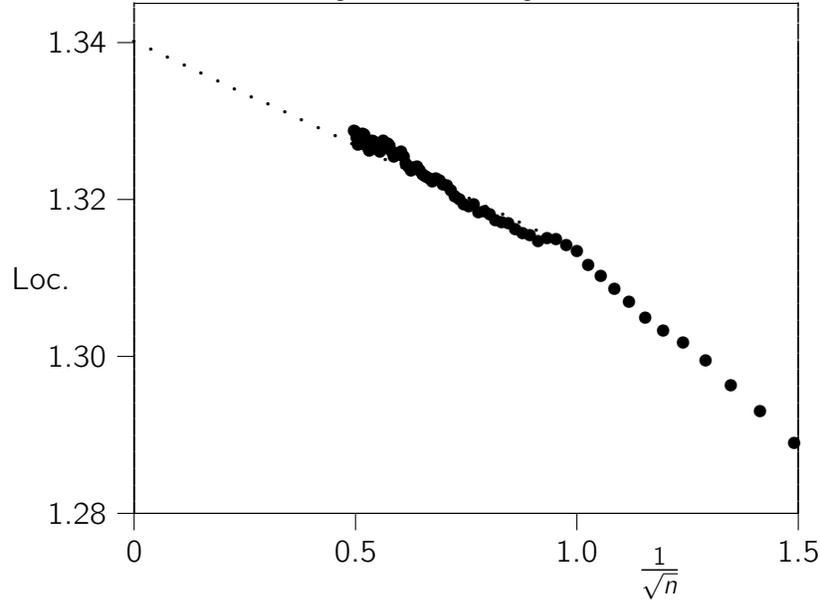

\input figure26-.tex
\caption{Extrapolating the intersections between specific heat curves
for  adsorbing walks in $\mathL_3^+$.  These data points are the locations
of intersections between specific heat curves for $n$ and for $n\plus 100$
(and for even $n$ and $n\in[46,400]$).  The points (except for
some points at the largest values of $n$) approximately line up
linearly if plotted against $1/\sqrt{n}$.  The location of the critical
point can be estimated by extrapolating the line to the left vertical axis;
this gives the estimate in equation \Ref{eqn29-3}.}
\label{figure3-3a}  %%ZXZ[figure3-3a]
\end{figure}

The locations of the intersections between the specific heat curves
in figure \ref{figure2-3a} is a function of $n$ and are estimates of the
critical adsorption point $a_c^+$.   By plotting the intersections 
between  $\C{C}_n(a)$ and $\C{C}_{n+100}(a)$ against
$\sfrac{1}{\sqrt{n}}$ (see figure \ref{figure3-3a}, where $n=2\, N$ 
and $N\in [23,200]$); it is seen that the intersections fall approximately
along a curve, which may be extrapolated to its intersection with the
vertical axis.  This gives a rough estimate of the critical point 
$\log a_c^+$ as being located in the the interval $[1.33,1.35]$.
A more accurate extrapolation is done by using a linear least squares model 
to extrapolate to $n=\infty$.  Fitting to the model $a_c^+\plus \sfrac{a_0}{\sqrt{n}}
\plus \sfrac{a_1}{n}$, for all $n \geq 50$, gives the estimate
$\log a_c^+ \approx 1.337$.  By examining the spread of the 
data in figure \ref{figure3-3a}, a confidence interval can be estimated:
\begin{equation}
\log a_c^+ = 1.337\pm 0.020 .
\label{eqn29-3}  %%ZXZ[eqn29-3]
\end{equation}
This estimate is slightly larger than the estimate in equation
\Ref{abest3}, but is consistent with the estimate $a_c^+ = 1.334$
in reference \cite{JvRR04}.  However, the noise in the data in figure
\ref{figure3-3a} makes this a less reliable estimate.

Equation \Ref{eqn27} is equally valid for adsorbing walks in the cubic
lattice.  The last term on the right hand side is equal to zero when $a=a_c^+$.   
Thus, by plotting $\log (\C{C}_n(a)/\C{C}_m(a))$ against
$\log (n/m)$, a set of curves should be seen which intersect when $a=a_c^+$.  
At this point the coefficient of $\log \sfrac{n}{m}$ is an estimate of $\alpha\phi$.
Since $\alpha=0$ in this model (and $\phi=\sfrac{1}{2}$), the critical point
can also be determined by solving for $a$ in
\begin{equation}
\log \LB \Sfrac{\C{C}_n(a)}{\C{C}_m(a)} \RB = 0 .
\end{equation}
Solving this for $n\in[150,500]$ and $m\in[n\minus 100,n\minus 10]$ gives
a large collection of estimates.  The average is
\begin{equation}
a_c^+ = 1.324 \pm 0.012,
\label{eqn27CC}  %%ZXZ[eqn27CC]
\end{equation}
where the confidence interval is one-half of the largest difference between
two estimates in the collection.  This result is smaller than the estimate in
equation \Ref{eqn29-3}, and larger than the best estimate in 
equation \Ref{abest3}.

These results indicate that there may remain sources of systematic errors in
the data and in the analysis, and that the estimates for $a_c^+$ should be
considered in this context.  
 
\begin{figure}[t]
\input figure27-.tex
\caption{The microcanonical density function $P^+(\eps)$ of visits 
for adsorbing walks in $\mathL_3^+$.  These data the finite size approximations 
$P_n^+(\eps)$ to $P^+(\eps)$ for $n=100$ and $n=500$, as well as the
extrapolated estimate of $P^+(\eps)$.  The right derivative of $\log P^+(\eps)$
at $\eps=0$ is an estimate of the location of the critical adsorption point
$a_c^+$ in the model.}
\label{figure9-3}  %%ZXZ[figure9-3]
\end{figure}

\subsubsection{The microcanonical density function:}
The microcanonical density function of visits in adsorbing positive walks is
determined from the microcanonical data in the model, and is given by
\begin{equation}
P^+(\eps) =\lim_{n\to\infty} (c_n^+(\lfl \eps n \rfl))^{1/n} = \lim_{n\to\infty} P_n^+(\eps),
\end{equation}
where $P_n^+(\eps) = (c_n^+(\lfl \eps n\rfl))^{1/n}$ is a finite size approximation 
to $P^+(a)$. Existence of $P^+(\eps)$ can be shown
(see for example reference \cite{JvR15}), 
and $\log P^+(\eps)$ is a concave function of $\eps$.  

$P^+(\eps)$ can be determined by interpolating the finite size approximations
$P_n^+(\eps)$ and then extrapolating to $n=\infty$ by fitting a least squares
model to the data.  In figure \ref{figure9-3} the data for the extrapolated
function $P^+(\eps)$ is plotted together with $P_n^+(\eps)$ for $n=100$ and
$n=500$. 

A least squares fit of a quadratic to $\log P^+(\epsilon)$ for $\eps\in[0,0.1]$ 
gives $\log P^+(\eps) \approx 1.54378\minus 0.28704\eps\minus 0.08784\eps^2$.
By taking the right derivative and then taking $\eps\to 0^+$, an
estimate for the critical point is obtained:
\begin{equation}
a_c^+ \approx 1.332 .
\label{eqn27EE}  %%ZXZ[eqn27EE]
\end{equation}
The free energy is the Legendre transform of $\log P^+ (\epsilon)$.  This
may be estimated by fitting a polynomial to $\log P^+(\epsilon)$.  If a cubic
polynomial in $\eps$ is fitted to $\log P^+(\eps)$ for $0 \leq \eps \leq 0.5$,
then the estimated free energy for $a>a_c^+$ is approximately
\begin{equation}
\fl
F (a) \approx  1.6526 -0.4240\log a  -  (0.3184 \minus 1.319\log a)
\sqrt{-0.2438 \plus 1.0105 \log a}.
\end{equation}
The critical point can be estimated as that location where the square root
in the above is zero.  This gives $a_c^+ \approx 1.273$.  Similarly, the
factor $(0.3184-1.319\log a)$ vanishes when $a_c^+ \approx 1.273$.
These estimates are far less secure than the estimates above, and are also
smaller. 

\begin{figure}[t]
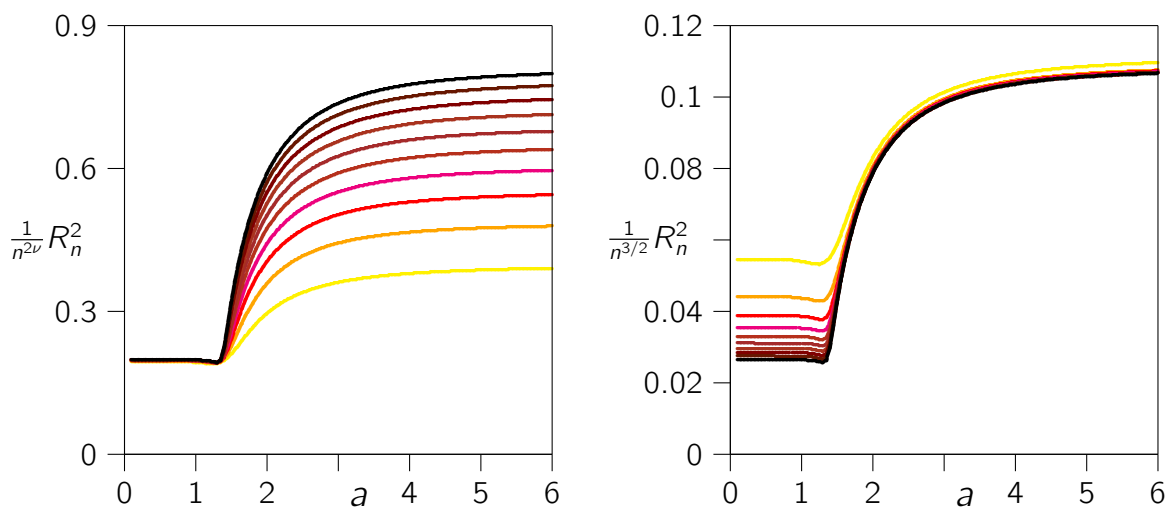

\input figure28-.tex
\caption{The mean square radius of gyration as a function of $a$
for  adsorbing walks in $\mathL_3^+$.  In
the left panel $R^2_n$ is divided by $n^{2\nu}$.  For $a<a_c^+$ this
shows that $R^2_n \sim n^{2\nu}$, and for $a>a_c^+$ $R^2_n$ 
increases in size faster than $n^{2\nu}$ (since it is adsorbed, it
stays in the vicinity of the adsorbing boundary, and it should
be the case that $R^2_n \sim n^{3/2}$ in this regime.  This is seen in
the right panel, where $R^2_n$ is divided by $n^{3/2}$.  For $a>a_c^+$
the curves collapse to a single underlying, exposing the scaling
in the adsorbed phase.  The lengths of the walks varied
from $n=50$ (yellow) with colours increasing in hue to black when
$n=500$ in steps of $50$. }
\label{figure8-3}  %%ZXZ[figure8-3]
\end{figure}

\begin{figure}[b]
\input figure29-.tex
\caption{The metric exponent as a function of $a$
for  adsorbing walks in $\mathL_3^+$.  The exponent
was estimated from equation \Ref{eqn41-3} by fixing $n$ and then
averaging over $m$ for $50\leq m \leq 500$.  With increasing
$n$ the change from $\nu=0.588\ldots$ in the desorbed phase, to
$\nu=0.75$ in the adsorbed phase will become a step function at
$a=a_c^+$.  The lengths of the walks varied
from $n=50$ (yellow) with colours increasing in hue to black when
$n=500$ in steps of $50$. }
\label{figure8-nu3}  %%ZXZ[figure8-nu3]
\end{figure}

\subsubsection{Metric data:}
The mean square radius of gyration $R_n^2$ and mean height $H_n$
of the endpoint of the walk can be calculated as a function of $a$.
In the desorbed phase (for $a<a_c^+$) it is expected that
$R_n^2 \sim n^{2\nu}$, and $H_n \sim n^\nu$, where $\nu = 0.587\ldots$
\cite{C10} is the metric exponent.  This scaling changes in the adsorbed phase
(when $a>a_c^+$); in this phase it should be the case that
$R_n^2 \sim n^{3/2}$ and $H_n \sim \hbox{constant}$, since
an adsorbed walk in the cubic lattice should have the statistics of a
walk in one dimension lower.

These expectations are supported by the data, as seen, for example, in figure 
\ref{figure8-3}, where data for the mean square radius of gyration are
plotted as a function of $a$.  These graphs clearly show two scaling
regimes, namely a high temperature phase (when $a< a_c^+$) where
the walk has bulk critical exponents and is desorbed, and a low 
temperature phase where the walk stays near the adsorbing boundary
and has critical exponents of a walk in one dimension lower.

The metric exponent $\nu$ may be estimated from 
$R_n^2$ by examining the ratios
\begin{equation}
2\,\nu_{n,m}(a) = \frac{\log ( R_n^2/R_m^2 )}{\log (n/m)}.
\label{eqn41-3}   %%ZXZ[eqn41-3]
\end{equation}
Here, $\nu_{n,m}(a)$ is a function of $n$ and $m$.   By averaging over
$m$, the estimate $\nu_n(a) = \LA \nu_{n,m}\RA_m$ can be determined.
Taking the average for $100\leq m \leq 500$ in multiples of $5$ 
(and for $m\not= n$) gives an estimate for $\nu_n(a)$.  The results are 
plotted in figure \ref{figure8-nu3} for $n\in\{50,100,150,\ldots,500\}$.
The data for $a\leq 1.2$ give $\nu \approx 0.592$, and for $a \geq 1.7$,
$\nu \approx 0.740$.

The scaling of $\nu_n(a)$ as a function of $\tau=n^\phi(a\minus a_c^+)$
can be uncovered by plotting the data in figure \ref{figure8-nu3}.  This
gives a set of curves which are very close to one another, uncovering a
scaling function \hbox{\Large$\nu$} where 
$\nu_n(a) =\hbox{\Large$\nu$}(\tau)$.

\begin{figure}[t]
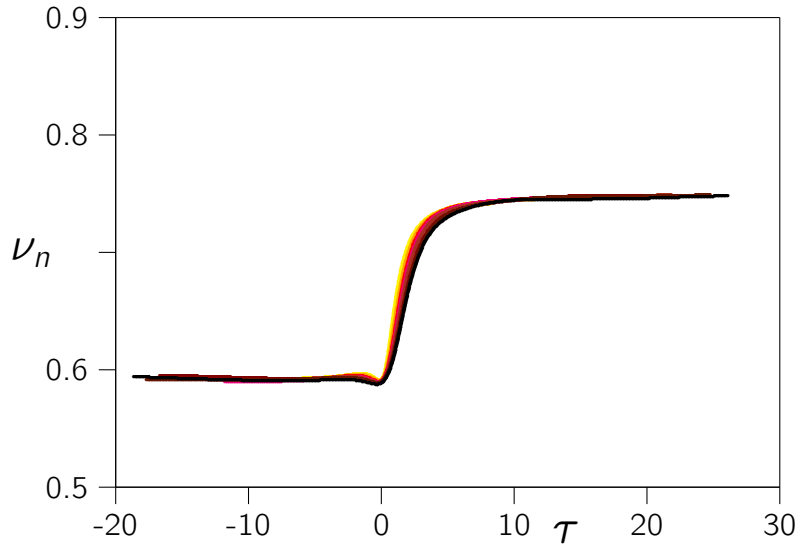

\input figure30-.tex
\caption{The metric exponent as a function of the rescaled
variable $\tau = n^\phi(a-a_c^+)$ for  adsorbing walks in $\mathL_3^+$.  
The lengths of the walks varied
from $n=50$ (yellow) with colours increasing in hue to black when
$n=500$ in steps of $50$. }
\label{figure6-D3}  %%ZXZ[figure6-D3]
\end{figure}

\begin{figure}[t]
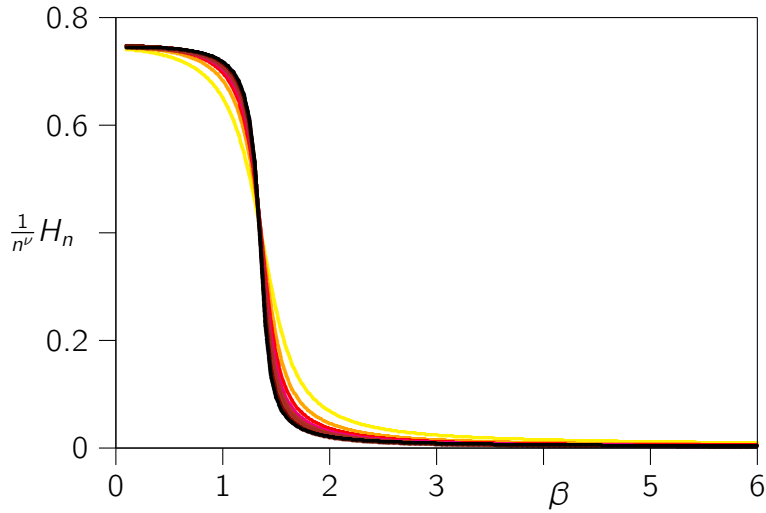

\input figure31-.tex
\caption{The mean height $H_n$ of the endpoint of adsorbing walks 
of length $n$ as a function of $a$ for  adsorbing walks in $\mathL_3^+$.  
On the left $n^{-\nu}H_n$ is plotted
as function of $a$.  This quantity decreases sharply as $a$ passes through
the adsorption critical point $a_n^+$.  This behaviour is also seen on
the right, where $n^{-1} H_n$ is plotted as a function $n$.  The lengths of the 
walks varied from $n=50$ (yellow) with colours increasing in hue to black when
$n=500$ in steps of $50$.}
\label{figure8HH3}  %%ZXZ[figure8HH3]
\end{figure}

The average height of the endpoint of the walk is plotted as a function
of $a$ in figure \ref{figure8HH3}.  The left panel displays the height
normalised by division with $n^\nu$ and gives a set of curves which increase 
with $n$ for $a< a_c^+$, and decrease with $n$ for $a>a_c^+$.  
The curves intersect close to the critical adsorption point, and the limiting curve 
(in the $n\to\infty$ limit) should be a step function with critical point at 
$a=a_c^+$. 

\begin{figure}[b]
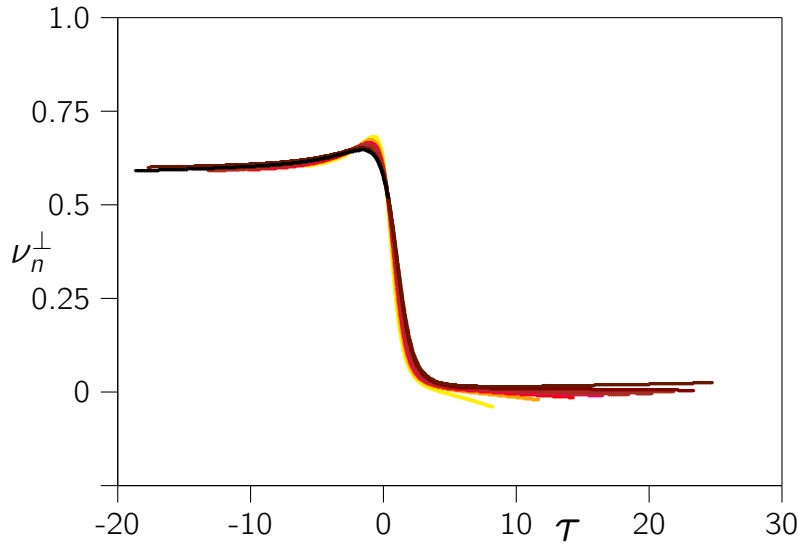

\input figure32-.tex
\caption{The vertical metric exponent $\nu^\perp$ as a function of $a$
for  adsorbing walks in $\mathL_3^+$.  
The exponent was estimated from equation \Ref{eqn42-3} by fixing $n$ and 
then averaging over $m$ for $100\leq m \leq 500$.  With increasing
$n$ the exponent changes from $\nu^\perp=\nu\approx 0.59$ in the desorbed 
phase, to $\nu^\perp=0$ in the adsorbed phase.  The lengths of the walks varied
from $n=50$ (yellow) with colours increasing in hue to black when
$n=500$ in steps of $50$.}
\label{figure6-nuH3}  %%ZXZ[figure6-nuH3]
\end{figure}

The vertical metric exponent $\nu^\perp$ can be estimated 
from $H_n$, by using a method similar to equation \Ref{eqn41-3}, namely
an approximation by examing the ratios of $H_n$:
\begin{equation}
\nu^{\perp}_{n,m} (a) = \frac{\log (H_n/H_m)}{\log (n/m)}.
\label{eqn42-3}   %%ZXZ[eqn42-3]
\end{equation}
The exponent is approximated by $\nu^{\perp}_n(a) = \LA \nu_{n,m}(a)\RA_n$.
Taking the average for $100\leq m \leq 500$ for fixed $n$ gives esimates 
for $\nu^\perp_n(a)$.  If $a<a_c^+$, then $\nu^\perp_n(a)$ should
have value approximately equal to $\nu$; that is, $\nu_n (a) \approx 0.58\ldots$, and 
for $a>a_c^+$, $H_n \simeq \hbox{const}$ so that $\nu_n^{\perp}(a) \approx 0$ 
in this phase.  The results are plotted in figure \ref{figure6-nuH3} against 
$\tau = n^\phi(a\minus a_c^+)$.

\begin{figure}[t]
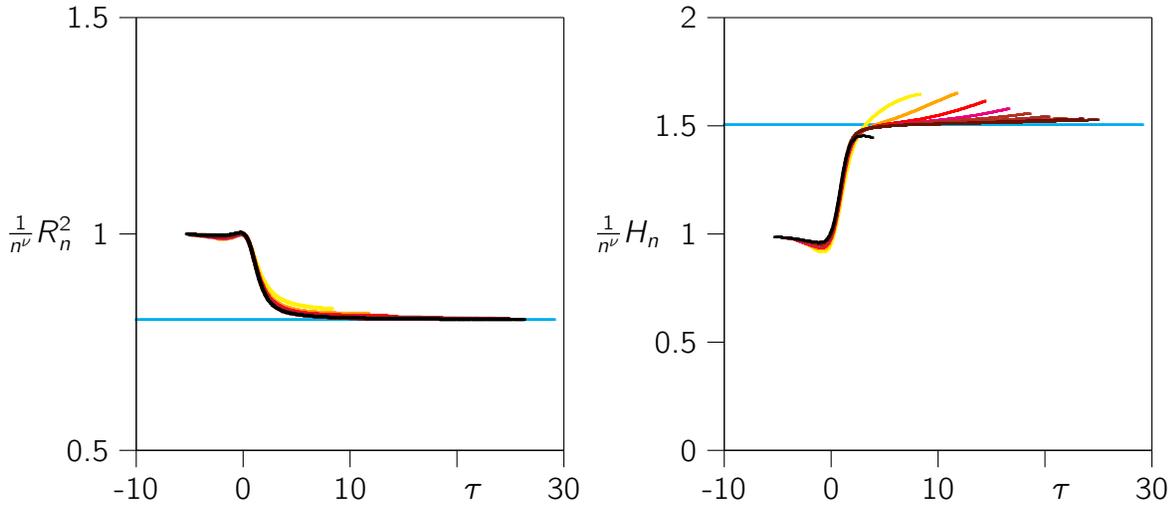

\input figure33-.tex
\caption{The rescaled mean square radius of gyration
$R^2_n/n^{2\nu_a}$ for  adsorbing walks in $\mathL_3^+$
plotted as a function of $\tau=n^\phi(a-a_c^+)$
is displayed on the left.  The metric exponent $\nu_a$ is the step-function
defined by $\nu_a=\frac{3}{4}$ if $a\leq a_c^+$ and $\nu_a = 1$ if 
$a>a_c^+$.  The rescaled curves collapse to a single underlying 
universal scaling function.  On the right the similar approach is used
to uncover the scaling function for the mean height of the endpoint of the
walk, scaled by $n^{\nu_a}$.  The data are curves for $n\in\{50,100,150,
\ldots,500\}$ increasing in hue from yellow ($n=50$) to black
($n=500$).}
\label{figure8RR3}  %%ZXZ[figure8RR3]
\end{figure}

Since $R^2_n \sim n^{2\nu_a}$ where $\nu_a = \nu \approx 0.588\ldots$
if $a<a_c^+$, and $\nu_a = \sfrac{3}{4}$ if $a>a_c^+$, the ratio of
$R^2_{2n}$ and $R^2_n$ is given by
\begin{equation}
\frac{R^2_{n}}{R^2_{2n}} \approx 2^{-2\nu_a} ,\q\hbox{or}\q
\frac{2^{2\nu}R^2_{n}}{R^2_n} \approx 
\cases{
1, & if $a< a_c^+$; \\
2^{2\nu - 2\nu_a} \approx 2^{-0.32}, & if $a>a_c^+$.
}
\end{equation}
In figure \ref{figure8RR3} this is plotted as a function of the rescaled
variable $\tau = n^{1/2}(a\minus a_c^+)$ for $n$ from $25$
to $250$ in steps of $25$.  The curves coincide well  with increasing $n$
and signals a transition when $a=a_c^+$ from the desorbed scaling regime into
the adsorbed scaling regime.   When $a>a_c^+$, $2^{-0.32} = 0.801\ldots$, as
shown in the graph.  A similar approach using the heights of
the endpoint would involve plotting
\begin{equation}
\frac{H_{n}}{H_{2n}} \approx 2^{-\nu_a} ,\q\hbox{or}\q
\frac{2^\nu H_{n}}{H_{2n}} \approx 
\cases{
1, & if $a< a_c^+$; \\
2^{\nu - \nu^\perp} \approx 2^{0.588}, & if $a>a_c^+$,
}
\end{equation}
where $\nu^\perp$ is the vertical metric exponent.  When $a>a_c^+$, 
$2^{0.588} = 1.503\ldots$, as shown in figure \ref{figure8RR3}.

\subsubsection{The generating function:}

\begin{figure}[t]
\centering
\includegraphics[scale=0.6]{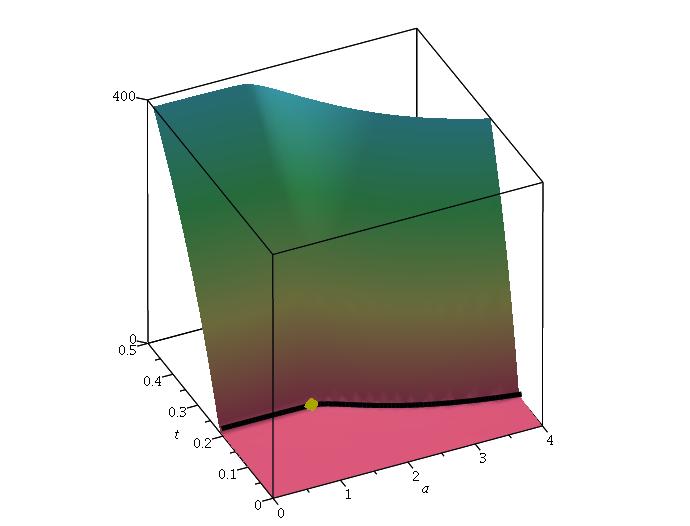}
\caption{$G_{500}(a,t)$ as a function of $(a,t)$
for  adsorbing walks in $\mathL_3^+$.  The critical
curve $t_c(a)$ is the black curve, and the location
of the critical point at $a=a_c^+$ is indicated.  Below the
critical curve $G(a,t)$ is finite, and approximated well by
$G_{N}(a,t)$ (for large $N$ and not too close to the critical
curve).  Above the critical curve $G(a,t)$ is divergent, while
$G_N(a,t)$ is finite.}
\label{figureAdsrb-3}   %%ZXZ[figureAdsrb-3]
\end{figure}

The generating function of adsorbing walks in the cubic lattice is given 
by equation \Ref{eqn45}, where $c_n^+(v)$ is again the number of walks
from the origin of length $n$ in $\mathL^3_+$, and with $v$ visits to the
adsorbing boundary $\partial\mathL^3_+$.  Approximations to
$G(a,t)$ are given by $G_N(a,t)$ in equation \Ref{eqn46a}, and 
$G_{500}(a,t)$ was calculated using the approximate values of
$c_n^+(v)$ obtained by sampling with the GAS algorithm.
The critical curve is given by equation \Ref{eqn46A}
(see figure  \ref{figure4-6}).  $G(a,t)$ is singular when $t=t_c^+(a)$, 
and if $t>t_c^+(a)$, then $G(a,t)$ is divergent.  In figure \ref{figureAdsrb-3}
the approximation $G_{500}(a,t)$ is plotted, with the location of the
critical curve, and critical point $(a_c^+,t_c^+)$ indicated
(where, as before, $t_c^+ = t_c^+(a_c^+)$).  The critical curve is
similar to the critical curve in figure \ref{figure4-6}, and the critical
point divides the critical curve into two curves, namely
a curve where the transition is a high temperature or desorbed walk
marked by $\tau_0$, and a curve where the transition is to a low
temperature or adsorbed walk marked by $\lambda$.
Along $\tau_0$ the critical curve is given by $t_c^+(a) = 
\sfrac{1}{\mu_3}$, for $a\leq a_c^+$ and where $\mu_3$ is
the growth constant of self-avoiding walks in the cubic lattice.

The critical curve is parametrized by the scaling fields 
$(\sigma,g)$ as shown in figure \ref{figure4-6}.   Here, the scaling
fields are given by $g = t_c^+\minus t$, and $\sigma=a\minus a_c^+$.
The singular points in $G(a,t)$ along $t_c^+(a)$ are described by the scaling
assumptions shown in equation \Ref{eqn46}.  The exponent $\gamma_1$ 
can be estimated by putting $a=1$ so that
$G(1,t) \sim g^{-\gamma_1}$.   By using the model in equation \Ref{eqn47},
the estimate
\begin{equation}
\gamma_1=0.725\ldots 
\end{equation}
is obtained.  This result is close to the estimate $0.697(2)$ in reference \cite{HG94}.
A similar analysis with $a=a_c^+$ gives the estimate
\begin{equation}
\gamma_s=1.203\ldots
\label{eqn71}   %%ZXZ[eqn71]
\end{equation}
for the surface exponent at the critical adsorption point.  This is slightly
smaller than the estimate $1.304(16)$ in reference \cite{ML88A}.

\begin{figure}[t]
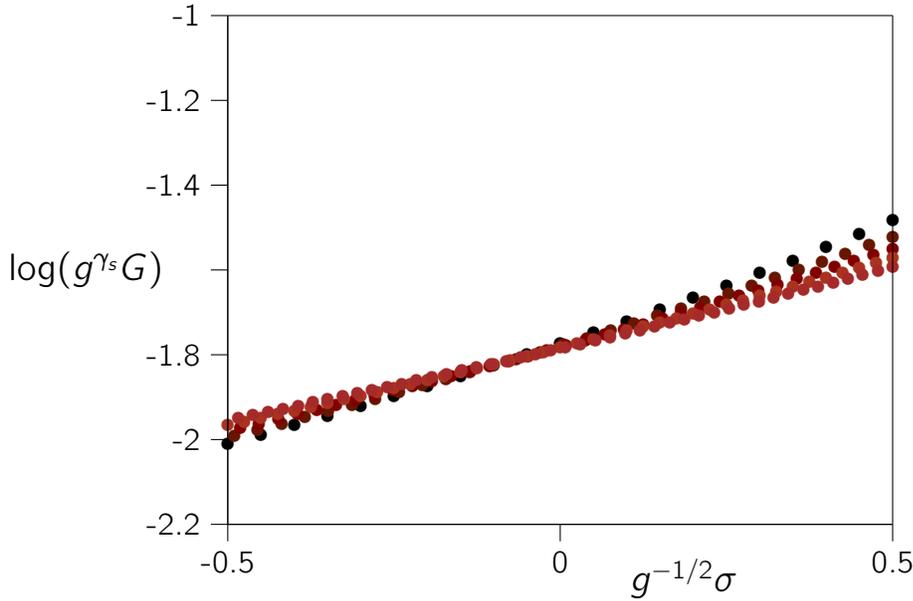

\input figure35-.tex
\caption{Scaling of the generating function near the critical adsorption point
for adsorbing walks in $\mathL^+_3$.
The data points are above were collected for $g\in[0.01,0.05]$ and
$\frac{1}{2}\sqrt{g} \leq \sigma \leq \frac{1}{2}\sqrt{g}$.  The value of the
exponent $\gamma_s$ was assumed to be given by the estimate in 
equation \Ref{eqn71g}.}
\label{figureGscale-3}  %%ZXZ[figureGscale-3]
\end{figure}

The situation is less clear in the adsorbed phase. The adsorbed walk 
should have the scaling of a self-avoiding walk in two dimensions,
so that $\gamma_+$ is given by the entropic exponent of walks 
in $d=2$ ($\gamma_+=\sfrac{43}{32}$).  Putting $a=4$ and plotting $Z_n^{1/n}(4)$ 
against $n$ gives an estimate for the critical value $t_c^+(4)$.  In this case 
the data quickly converges to $t_c^+(4)=0.0926\ldots$ (to four decimal places).  
Assuming that $t_c^+(4)=0.0926$ and choosing the scaling field 
$g=(0.0926 \minus t)$ gives a model similar to equation \Ref{eqn48}.  
Plotting the left hand side as a function of $\sfrac{1}{\log g}$, 
and fitting it to a quadratic for $t\in[0,0.06]$, gives the estimate 
$\gamma_+ \approx 1.2\ldots$, still well below the expected result 
$\sfrac{43}{32} = 1.34375$.  Examination of the data shows strong dependence 
of this result on the range of $g$ in the model.  For example, a fit 
with $t\in[0,0.09]$ gives a smaller value $\gamma_+ \approx 1.1\ldots$.
These variable results indicate that $G_{500}(a,t)$ is not a 
good approximation to $G(a,t)$ near the critical curve $\lambda$ 
in figure \ref{figure4-6} for adsorbing walks in the cubic lattice.

In the vicinity of the critical point $(a_c^+,t_c^+)$ the generating function
should exhibit scaling given by equation \Ref{eqnGscale}.
Plotting $g^{\gamma_s}\,G(a,t)$ (with $\gamma_s=1.203$) 
against the combined variable $g^{-1/2}\sigma$ should
expose the scaling function $f$. In figure \ref{figureGscale-3} this 
is done by plotting $\log (g^{\gamma_s}\,G(a,t) )$ against $g^{-1/2}\sigma$ for 
$g=\sfrac{1}{\mu_3}\minus t \in [0.01,0.05]$ and $\sigma
=(a\minus a_c^+) \in [-\sfrac{1}{2}g^{1/2},\sfrac{1}{2}g^{1/2}]$.

\begin{figure}[t]
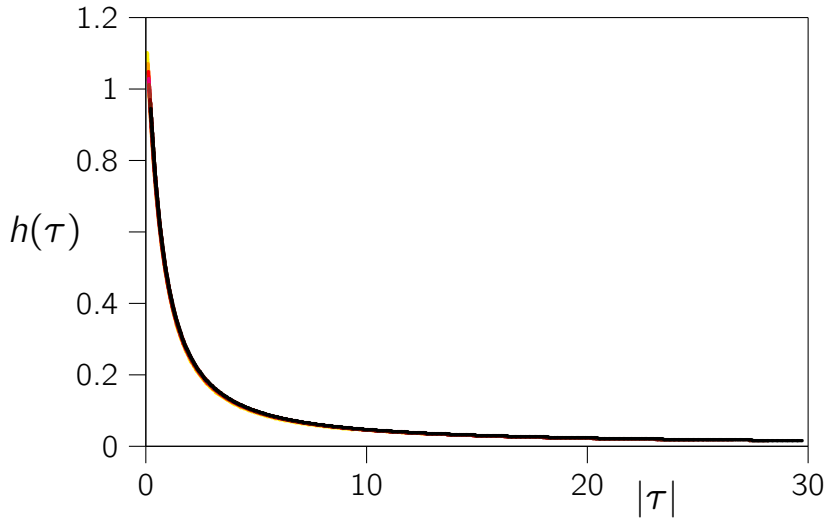

\input figure36-.tex
\caption{The scaling of the partition function as suggested by equation
\Ref{eqn73} for adsorbing walks in $\mathL^+_3$.  
The curves are plots of $n^{-0.203} Z_n(a)\mu^{-n}$ as a function
of $|\tau | = n^{1/2} |a\minus a_c^+|$ (with $a_c^+$ approximated by
$1.33$ and $a<a_c^+$).  All the data collapse to a single curve, which 
is the scaling function $h$
in equation \Ref{eqn73}.  The lengths of the walks varied
from $n=50$ (yellow) with colours increasing in hue to black when
$n=500$ in steps of $50$. }
\label{figureZscale3}  %%ZXZ[figureZscale3]
\end{figure}

The partition function (see equation \Ref{eqnZ}) also exhibit scaling for 
large $n$.  The scaling assumption in equation \Ref{eqn53} applies
here as well,  where $\log \mu_a = \C{F}(a)$ and $\log \mu_a = - \log t_c(a)$.
As before, the exponent $\gamma_t$ is related to the $\gamma_s$-exponent
as in equation \Ref{eqn71g}:
\begin{equation}
\gamma_t-1 = \gamma_s - 1 \approx 1.203 - 1 = 0.203.
\end{equation}
That is, when $a$ is close to $a_c^+$ (and $a<a_c^+$), then the partition 
function has asymptotic behaviour
\begin{equation}
Z_n(a) \simeq n^{0.203} h(n^\phi (a_c^+\minus a))\, \mu_a^n .
\label{eqn73}   %%ZXZ[eqn73]
\end{equation}
This result may be tested by plotting $n^{0.203} Z_n(a_c^+) \, (t_c(a))^n$ 
against $\tau = n^\phi (a\minus a_c^+)$.  This scaling is seen in figure
\ref{figureZscale3} for $a<a_c^+$ and $n\in\{50,100,150,\ldots,500\}$; 
all the data accumulate along a single curve,
exposing the scaling function $h$.

\section{Conclusions}
\label{section4}  %%ZXZ[section4]

The adsorbing self-avoiding
walk is a classical model in rigorous and numerical statistical mechanics, 
and have received considerable attention in the physics and mathematics
literature \cite{deG79,HTW82,LM88A,HG94,JvRR04}.  

In this paper the feasibility of collecting data in the microcanonical ensemble
on adsorbing walks using a flat histogram implementation of the GAS algorithm
\cite{JvRR09} was considered.  This is an approximate enumeration algorithm, and the
data can be used to directly estimate partition and generating functions, from
which thermodynamic functions such as the free energy and specific heat
can be determined (see, for example, reference \cite{JvR15}). 

The implemementation of the algorithm was done using endpoint elementary
moves on half space self-avoiding walks, and the algorithm sampled from
a flat histogram with reasonable success in both length and energy in
the square and cubic lattices.  Analysis of the data gives good results, 
better than previous Monte Carlo simulations in, for example,
references \cite{HG94,JvRR04}, but not as good as exact enumeration data
in references \cite{BWO99,BGJ12}.  A significant advance of this algorithm is that
its produces a large amount of microcanonical data. Modifications to 
obtained data with respect to other quantities are trivial, and can
easily be implemented.  The simulations reported here were done on
a Dell Inspiron 530 desktop machine, but note that the algorithm can
be implemented in parallel on a cluster with each cluster generating
an independent sequence.  This should give radically improved
statistical data.

The success of the implementation
suggests that this numerical method may be used on other models
(collapsing self-avoiding walks \cite{ML89,TJvROW96}, for example). 
However, it may be necessary to extend the method by introducing, in
addition to the sets of parameters denoted by $\{\beta_{\ell,u}\}$ and
$\{\gamma_{\ell,u}\}$, additional sets of parameters which are
conjugate to classes of elementary moves.  For example, in the model
of collapsing walks (see figure \ref{figure3}), the energy may be changed
by $\Delta u \in \{-2d\plus 1,-2d\plus 2, \ldots,2d\minus 1\}$ by an
elementary move, and parameters may be introduced for each value of 
$\Delta u$ to achieve flat histogram sampling (in a way similar to the
introduction of $\gamma_{\ell,u}$ for elementary moves increasing
the energy of the walk. This will increase the complexity of the implementation, 
but with the result that flat histogram sampling will be easier to achieve. 

In the models of square and cubic lattice adsorbing walks, the algorithm
produced data which gave good estimates of the locations of the critical
adsorption point.  The best estimates are obtained from equations \Ref{abest}
and \Ref{abest3}, namely
\begin{equation}
a_c^+ =
\cases{
1.779 \pm 0.003, & \hbox{in the square lattice}; \\
1.306 \pm 0.007, & \hbox{in the cubic lattice}.
}
\end{equation} 
These results can be used to estimate the crossover exponent $\phi$
association with the adsorption transition, and our best esimates
are seen in equations \Ref{phibest} and \Ref{phibest3}:
\begin{equation}
\phi =
\cases{
0.496 \pm 0.009, & \hbox{in two dimensions}; \\
0.505 \pm 0.006, & \hbox{in three dimensions}.
}
\end{equation} 
In addition, other quantities from which $a_c^+$ and $\phi$ can be estimated
were examined, and results largely consistent with the above values were
obtained (see, for example, equations \Ref{eqn29}, \Ref{eqn27AA}  and
\Ref{eqn27BB} for square lattice results, and equations \Ref{eqn29-3},
\Ref{eqn27CC} and \Ref{eqn27EE} for cubic lattice results).  These
numerical estimates are in good agreement with those presented
in reference \cite{JvRR04}, and also in reference \cite{BGJ12} in the
case of the square lattice.  The estimate in this reference, obtained
from exact series data, namely $a_c^+ = 1.77564$, agrees with the 
estimate above to two decimal places.  In the cubic lattice the estimate
for $a_c^+$ above is slightly smaller than the estimates $a_c^+=1.338
\pm 0.005$ \cite{ML88A} and $a_c^+ = 1.334 \pm 0.027$ in reference
\cite{JvRR04} (rounding up of this last error bar gives a confidence 
interval which includes $1.306$).

The signature of the adsorption transition in the metric quantities of the
model was also examined.  The scaling of these quantities with
$\tau = n^{\phi}(a\minus a_c^+)$ were plotted in figures \ref{figure6-D},
\ref{figure6-nuH} and \ref{figure8RR} in the square lattice, and
in figures \ref{figure6-D3}, \ref{figure6-nuH3} and \ref{figure8RR3}
in the cubic lattice.  These results show a transition strongly characterised
by changes in metric scaling and verify the value of the metric exponent
and its finite size scaling through the critical point.

Finally, the scaling of the generating function and partition function
in these models were examined.  Our results strongly supports the
conventional properties of the model, and the values of the
exponents $\{\gamma_1,\gamma_s,\gamma_+\}$ estimated here
are consistent with exact values and other estimates in the literature.

The results in the square lattice are consistent with the exact values of
$\gamma_1$ and the surface exponent $\gamma_s$, and the generating
and partition partition function exhibit scaling consistent with the value
of $\gamma_s$, as shown in figures \ref{figureGscale} and \ref{figureZscale}.
In the cubic lattice our data gave the estimates $\gamma_1 \approx 0.725$
and $\gamma_s \approx 1.203$.  These values are in addition to estimates
elsewhere in the literature (see references \cite{HG94}, \cite{ML88A}), and although
the estimates here may be improvements on previous esimates, they remain
uncertain.  However, scaling of the generating function in figure
\Ref{figureGscale-3}, and of the partition function in figure \ref{figureZscale3},
is some evidence that the esimate for $\gamma_s$ is at least consistent
with the scaling in the model.

\vspace{1cm}
\noindent{\bf Acknowledgements:} EJJvR acknowledges financial support 
from NSERC (Canada) in the form of a Discovery Grant.  

\vspace{1cm}
\noindent{\bf References}
\bibliographystyle{plain}
\bibliography{References}

\end{document}